\documentclass[twocolumn]{aastex62}

\usepackage{natbib}
\bibpunct{(}{)}{;}{a}{}{,} 
\usepackage{graphicx}
\usepackage{txfonts}
\usepackage{ulem}

\newcommand{\xmm}{{\textit{XMM-Newton}}}

\shorttitle{Green Peas in X-Rays}
\shortauthors{Svoboda et al.}

\begin{document}

\title{Green Peas in X-Rays\footnote{Based on observations obtained with {\xmm}, an ESA science mission with instruments and contributions directly funded by ESA Member States and NASA.}}

\correspondingauthor{Ji\v{r}\'{i} Svoboda}
\email{jiri.svoboda@asu.cas.cz}

\author{J.~Svoboda}
\affiliation{Astronomical Institute, Academy of Sciences, Bo\v{c}n\'{\i}~II~1401, CZ-14131~Prague, Czech~Republic}
\author{V.~Douna}
\affiliation{Instituto de Astronom\'{i}a y F\'{i}sica del Espacio, CONICET-UBA, Casilla de Correo 67 Suc. 28, C1428ZAA~Ciudad Aut\'{o}noma de Buenos Aires, Argentina}
\author{I.~Orlitov\'{a}}
\affiliation{Astronomical Institute, Academy of Sciences, Bo\v{c}n\'{\i}~II~1401, CZ-14131~Prague, Czech~Republic}
\author{M.~Ehle} 
\affiliation{ESA, European Space Astronomy Centre, Apdo. de Correos 78, Villanueva de la Ca\~{n}ada, E-28691 Madrid, Spain}

\begin{abstract}

Green Peas represent a population of compact, highly star-forming dwarf galaxies at redshifts $z\sim0.2-0.3$ that have recently been found to show signatures of ultraviolet ionizing radiation leakage. They are being considered as analogs to high-redshift star-forming galaxies, possibly responsible for the cosmic reionization. Despite the intensive studies of Green Peas in the ultraviolet and optical domains,
their X-ray properties have only so far been probed by nearby analogs.
In this paper, we present the first measurements of Green Peas in the X-ray domain to constrain their spectral properties and fluxes at high energies.
We analyzed {\xmm} observations of three Green Pea sources. For two of them, we found an X-ray luminosity exceeding by a half-order of magnitude its predicted value, derived from the star formation rate and metallicity.
Only an upper limit of the X-ray luminosity was derived for the third studied galaxy.
Our results thus indicate that at least some Green Peas  
produce copious amounts of highly energetic photons, larger than detected in other star-forming galaxies.
We discuss possible physical scenarios for the measured X-ray excess, including the presence of a hidden active galactic nucleus, a larger population of X-ray binaries, or ultra-luminous X-ray sources. 
Future spatially resolved X-ray images will discriminate between the models. 
Larger Green Pea samples
will provide a possible link between the X-ray properties and the leaking ultraviolet radiation.

\end{abstract}



\section{Introduction} \label{sec:intro}

Cosmic reionization, which took place between redshifts $z\sim20$ and $z\sim6$ \citep[e.g.][]{Robertson2015, Bowman2018}, 
was a pivotal change in the history of the universe,
affecting the galaxy formation and observability 
\citep{Gunn1965, Barkana2001, Choudhury2006, Zaroubi2012}. Discussions 
about the astrophysical sources that were responsible for this change
are still ongoing, and solid observational evidence is only now emerging. 
%

Hydrogen ionization is efficiently achieved by  
ultraviolet (UV) Lyman-continuum (LyC) radiation at wavelengths $\lambda\lesssim\!912$\,\AA,  produced by two main astrophysical sources: star-forming galaxies and quasars
\citep[e.g.][]{FaucherGiguere2008,Robertson2010,Bouwens2015,Giallongo2015}.
Higher-energy photons such as X-rays from high-mass X-ray binaries \citep[e.g.][]{Mirabel2011,Fragos2013,Jeon2014,Xu2014,Artale2015,Madau2017,Sazonov2018} and radiation from dark matter annihilation/decay processes \citep[e.g.][]{Mapelli2006,Valdes2007,Liu2016b} have been proposed to have contributed to the ionization and to the heating of the intergalactic medium (IGM) at large scales. The contribution of cosmic rays has also been considered, either those accelerated by the first supernovae \citep[e.g.][]{Nath1993,Sazonov2015,Leite2017} or in the jets of high-redshift microquasars \citep[e.g.][]{Tueros2014,Douna2017}.

Reionization by LyC photons poses two major questions: which LyC sources were dominant and whether their LyC production was sufficient.  
While quasars are powerful LyC producers, 
they may have been
%
%
too rare at $z>4$  \citep{Fontanot2012, Haardt2015}. Nevertheless, 
recent observational updates 
opened again the possibility of their dominant role
\citep{Giallongo2015,Madau2015}.
In contrast, star-forming galaxies 
were abundant in the young universe \citep{Robertson2010, Stark2016},
and could have been natural sources of ionizing UV photons produced by young massive stars. However, much of this radiation was probably consumed by the interstellar medium (ISM), and so the role of galaxies in the IGM ionization has not been resolved.

Sizeable galaxy samples (thousands of targets) are 
now available at redshifts $z>2,$ reaching out to $z\sim10$ \citep[e.g.][]{Ouchi09b, Ouchi18, Schenker2013,Bouwens15,Robertson2015,Hashimoto18,Oesch18}. 
Their direct observation 
in the LyC is impossible for $z>6$ 
due to large amounts 
of intergalactic neutral hydrogen. Therefore, we have to rely on observations at lower redshifts and theoretical modeling.
Galaxies with a sufficiently large LyC escape fraction have been detected only recently:
four targets at $z = 2-4$
\citep{deBarros16,Shapley2016,Bian2017,Vanzella18}, and 11 targets at $z\!\sim\!0.3$ \citep{Izotov2016,Izotov2016b,Izotov2018a,Izotov2018b}.  Their mean LyC escape fraction is $\sim\!20$\%. 
According to numerical simulations, 
an average escape of 20\% would be necessary to reionize the IGM at $z\!>\!6$ \citep{Yajima09, Paardekooper2015}.

The LyC-leaking galaxies at $z\!\sim\!0.3$ belong to (or are similar to) the population known as the Green Peas (GPs). 
The name Green Peas was originally coined for galaxies studied by \citet{Cardamone2009} at $z\sim0.2$ using
the optical Sloan Digital Sky Survey (SDSS). An extension to 
other redshifts was then proposed by \citet{Izotov2011}. GPs are compact, low-mass ($\sim\!10^{9}\, M_{\odot}$), 
and highly star-forming ($\sim\!10\, M_{\odot}$ yr$^{-1}$) 
galaxies with 
sub-solar metallicities (12+$\log$[O/H] $<8.7$).
They produce strong optical and UV emission lines that are formed by reprocessing of the ionizing radiation of massive stars: the equivalent widths (EWs) of their H$\alpha$ and [O~{\sc iii}]\,5007 lines can reach EW$\sim\!10^3$\,\AA, which makes them one-to-two orders larger than the typical EWs in SDSS galaxies, and match those typically encountered at $z>2$ \citep{Schaerer16}.

Furthermore, GPs have strong UV Lyman-$\alpha$ lines \citep{Henry15,Verhamme2017,Orlitova2018}, comparable to high-$z$ starburst galaxies known as the Lyman-Alpha Emitters (LAEs). Lyman-$\alpha$ is an optically thick line in galactic conditions, and is therefore a sensitive
tracer of the H\,{\sc i} parameters, including its velocities and column densities. The similarities between Lyman-$\alpha$ in GPs and LAEs suggest similar ISM conditions and a possible LyC escape in some LAEs.   
The unique features of GPs 
and their resemblance to high-redshift galaxies reinforce the need to study their properties and the origin of their ionizing and energetic emission in detail.

In this paper, we focus on the X-ray properties of GPs. 
X-rays mainly probe the presence of active galactic nuclei (AGNs) and accreting compact objects, which produce high-energy photons and particles. Consequently, they provide a radiative and mechanical 
feedback to the gas
in their surroundings, affecting its ionization and thermal state.
These sources may thus be responsible for creating ISM conditions that lead to the LyC escape. 
The analysis of highly ionized galaxies and galaxies with a LyC leakage 
will allow us to check whether their X-ray luminosities 
follow the trends established for other star-forming galaxies, or if additional sources such as AGNs need to be considered, which would affect the interpretation of the LyC sources now and at the era of reionization. 

The X-ray emission of star-forming galaxies has been shown to scale with the star-formation rate 
\citep[SFR;][]{Grimm2003,Ranalli2003,Mineo2012,Mineo2014}. 
The X-ray output from recent stellar populations
is dominated by high-mass X-ray binaries (HMXBs) whose lifetimes are of the order of $10^7$\,yr and should thus closely follow the star-formation events \citep[e.g.][]{Fabbiano06}. 
Low-mass X-ray binaries (LMXBs) have longer lifetimes and their X-ray emission scales with the galaxy mass. Their contribution to the observed X-ray luminosity can often be neglected in galaxies
with a high level of star formation per unit stellar mass \citep{Mineo2012b}. 
In contrast, the contribution of hot gas, powered by the massive-star feedback, hence proportional to the SFR, 
can be non-negligible \citep{Mineo2012b}. 
Recent studies have established an additional dependence of the
X-ray luminosity on metallicity 
\citep{Mapelli2010, Kaaret2011, Basu-Zych2013,Basu-Zych2016,Brorby2014,Brorby2016,Douna2015}. 
Metallicity modifies the cooling rates in the ISM, affects the number of HMXB systems, and may also modify the luminosity of the individual HMXBs \citep[e.g.][ and references therein]{Mirabel2011,Douna2015,Basu-Zych2016}. 
Translated to the Green Pea galaxies, large X-ray luminosities can be expected, due to their large SFRs and low metallicities.

 
We have obtained the first X-ray observations of GPs, using the {\xmm} satellite. We describe the observations and the data reduction in Section~\ref{method}. Results of the X-ray analysis and their presentation in the context of multi-wavelength data and a comparison with other galaxy samples are given in Section~\ref{results}. 
We discuss the results with respect to the previously published works in Section~\ref{Discussion}.
The main conclusions are summarized in Section~\ref{Conclusions}.

We used the most recent measurements of the cosmological parameters: $H_0 = 67.8$\,km\,s$^{-1}$\,Mpc$^{-1}$, $\Omega_M=0.308$, $\Lambda_0=0.69$ \citep{PlanckCollaboration2015} for the spectral model fits
and for the derivation of the X-ray luminosity from the observed flux.
$L_{\rm x}$ denotes the luminosity in the X-ray 0.5-8\,keV band, if not stated otherwise.

\begin{table*}[tb!]

\caption{{\xmm} observations of three Green Pea galaxies}
\centering
\begin{tabular}{ccccccc}
 \rule{0cm}{0.5cm}
 Source & R.A. (J2000) & Dec. (J2000) &  Redshift & Observation ID & Observation Date & Net Exposure Time [ks] \\
 \hline \hline
 \rule[-0.7em]{0pt}{2em} 
SDSSJ074936.7+333716 (GP1) & 117.403215 & 33.621219 &  0.2733   & 0690470101 &  2013 Mar 25    & 35.0 \\
 \rule[-0.7em]{0pt}{2em}   
SDSSJ082247.6+224144 (GP2) & 125.698590 & 22.695578 &  0.2162   & 0690470201 &  2013 Apr 6   &  31.5  \\
 \rule[-0.7em]{0pt}{2em} 
SDSSJ133928.3+151642 (GP3) & 204.867933 & 15.278369 &  0.1920    &  0690470401 &  2013 Jan 19   &  25.5 \\
\end{tabular}
\tablecomments{
Redshifts were adopted from the SDSS archive. 
}
\bigskip
\label{observations} 
\end{table*}

 \begin{table}[tb!]

 \caption{Physical Properties of Green Peas Studied in This paper.}
 \centering
 \begin{tabular}{ccccc}
  \rule{0cm}{0.5cm}
   & SFR$^{a}$ &  $\log M_*^{b}$   & $\log(\rm sSFR)$  & $\log[O/H]+12^{c}$\\
   & ($M_{\odot}$\,yr$^{-1}$) & ($M_\odot$) & (yr$^{-1}$) & \\
  \hline \hline
  \rule[-0.7em]{0pt}{2em} 
 GP1    &  $58.8$    & $9.8$ &  $-8.0$ &  $8.3$ \\
  \rule[-0.7em]{0pt}{2em}   
 GP2    &  $37.4$  &  $9.6$  &  $-8.0$ &  $8.1$  \\
  \rule[-0.7em]{0pt}{2em} 
 GP3    &  $18.8$  &  $9.3$  &  $-8.3$ &  $8.1$ \\
 \end{tabular}
 \tablecomments{
 $^{a}$ \citet{Cardamone2009} \\
 $^{b}$ \citet{Brinchmann2004} \\ 
 $^{c}$ This work.
 }
\label{SFR_table} 
\end{table}

\begin{figure}[tb!]
 \includegraphics[width=0.49\textwidth]{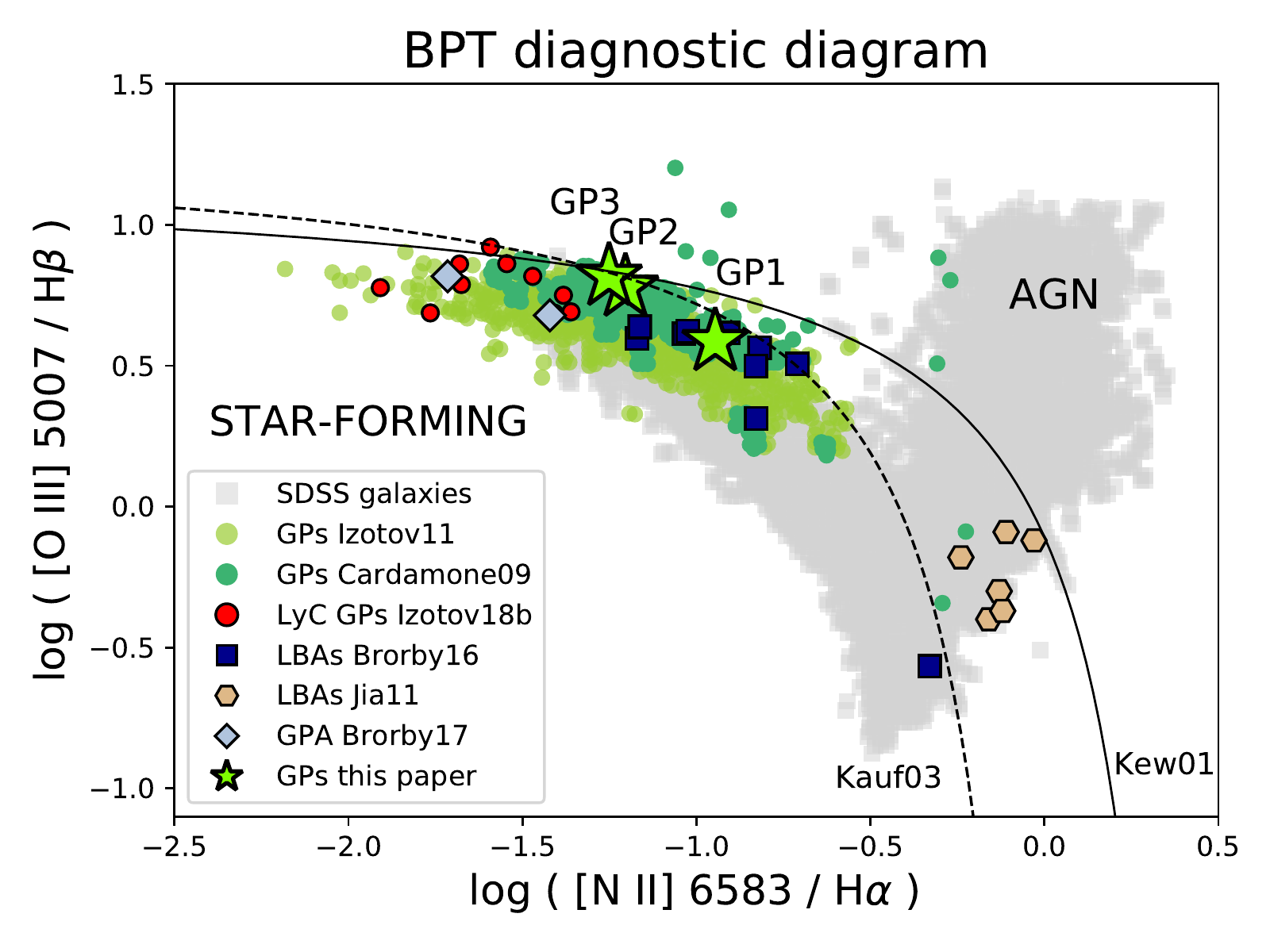}
 \caption{BPT diagram for emission-line galaxy classification. We present here the three GP targets that we observed with {\xmm}, together with other relevant samples discussed in this paper: the parent Green Pea sample from \citet{Cardamone2009}, its extension from \citet{Izotov2011}, Green Peas with escaping Lyman continuum \citep{Izotov2018b}, 
 Lyman-break analogs \citep{Jia2011, Brorby2016}, and Green Pea Analogs \citep{Brorby2017}. We also show the location of the bulk of the SDSS galaxies and the model lines of \citet{Kewley2001} and \citet{Kauffmann2003} that separate star-forming galaxies from AGN. 
 }
 \label{fig_bpt}
\end{figure}

\section{Observations and data reduction}\label{method}

\subsection{Sample Selection} \label{sample}
 

We selected GPs for the X-ray observations from \citet{Cardamone2009}. 
The parent sample consists of 80 star-forming galaxies with physical properties published in their Table~4 \citep{Cardamone2009}, and 20 targets that were interpreted as either Seyfert galaxies or transition objects with a mixed contribution from AGNs and star-formation. The classification was based on the ``BPT'' diagram \citep{Baldwin1981}, which uses the ratios of strong optical emission lines, 
[O\,{\sc iii}]\,$\lambda$5007/H$\beta$ and [N\,{\sc ii}]\,$\lambda$6583/H$\alpha.$ 

From this parent sample,
we selected targets with the largest SFR among the GPs that were classified as purely star-forming galaxies.
The X-ray emission of galaxies is expected to correlate with 
the SFR, therefore this selection maximizes the chances of X-ray detection. The sample was then narrowed down to three targets due to the {\xmm} exposure time constraints. Our pilot GP sample is therefore composed of three large-SFR targets that had the lowest redshifts and the most favorable {\xmm} observability. 
We list the \xmm-observed targets in Table~\ref{observations} and we further refer to them as GP1, GP2, and GP3. Their physical properties are listed in Table~\ref{SFR_table}.
The targets lie in the redshift range $z=0.19 - 0.27,$ their star formation rates SFR\,$\sim\!20-60\,M_{\odot}$\,yr$^{-1}$ 
\citep[based on H$\alpha$ measurements from][]{Cardamone2009},
their stellar masses $M_* \sim\!10^{9.6-9.8}$\,$M_{\odot}$ \citep[from][]{Brinchmann2004}, and their metallicities $\log[O/H]+12\sim8.1-8.3$ 
\citep[measured here using the prescription of][]{Pettini2004}.
Details about the data analysis are provided in Section~\ref{optical_analysis}.

With respect to the parent sample, the three selected targets have masses, SFRs and metallicities slightly above average (median GP M\,$_*\!\sim10^{9.5}$\,$M_{\odot},$
median SFR $\sim\!10M_{\odot}$\,yr$^{-1},$ median $\log[O/H]+12\sim8.07$).\footnote{We here considered metallicities derived by \citet{Izotov2011} for the \citet{Cardamone2009} GP sample. This was for consistency with 
our analysis methods that we describe in Section~\ref{optical_analysis}.} 
\citet{Izotov2011} proposed an extension of the Green Pea sample to a larger redshift range $z= 0.02-0.63,$ generalizing the color selection criteria. They obtained a sample of $\sim\!800$ GP-like galaxies with the median $M_*\sim10^{9}$\,$M_{\odot},$ median 
SFR\,$\sim\!4\:$\,$M_{\odot}$\,yr$^{-1},$ and median metallicity $\log[O/H]+12\sim8.1.$  GP1 has one of the largest SFRs in both the parent and the extended samples.

In Figure~\ref{fig_bpt}, we present the BPT diagram for the selected GP targets, together with the parent sample from \citet{Cardamone2009} and the extended sample from \citet{Izotov2011}.  
We use a different symbol (red dot) for the GP-like galaxies with a confirmed LyC escape \citep{Izotov2018b} because of their general interest for this paper. 
We represent several other galaxy samples that are pertinent to our GP study (see the next paragraph), and we show the location of the bulk of the SDSS galaxies. 
Finally, we plot the two most commonly used theoretical/empirical lines that
separate star-forming galaxies from AGNs \citep{Kewley2001, Kauffmann2003}. 
GP1, GP2 and GP3 lie in the star-forming section of the diagram, close to the separation line. GP2 and GP3 are in the middle of the GP cloud and on the Kewley line, while GP1 is more offset, due to its slightly higher metallicity. 

\subsection{Control Sample} 

For the comparison of GP X-ray luminosities with other star-forming galaxies,
we defined a control sample of X-ray-observed star-forming galaxies based on \citet{Douna2015}, \citet{Brorby2016}, and \citet{Brorby2017}. 
The sample of \citet{Douna2015} comprises $\sim\!\!30$ nearby ($z\!<\!0.02$) 
star-forming galaxies selected from \citet{Mineo2012}, 
representing a wide range of masses, metallicities, and SFRs.
The X-ray sources (X-ray binaries) have been 
individually resolved in these galaxies. 
We then included two metal-poor galaxies studied by \citet{Brorby2017}, who named them Green Pea analogs (GPAs) and selected them by similar criteria as \citet{Izotov2011}, but restrained the redshift to $z<0.1.$ 
Another interesting sample was provided by 
\citet{Brorby2016}, who focused on Lyman-break analogs (LBAs), 
building on previous results of \citet{Basu-Zych2013}. LBAs were defined by \citet{Heckman2011} using their far-ultraviolet (FUV) luminosity, mass, and metallicity so as to resemble high-$z$ Lyman-Break Galaxies (LBGs).
As the UV selection alone can lead to the presence of AGN in the sample, \citet{Brorby2016} cleared their LBA sample of targets with AGN signatures, and kept only pure star-forming galaxies.
Furthermore, as GPs are considered excellent analogs of high-redshift galaxies, we added a stacked sample of $z\!\sim\!2$ LBGs from \citet{Basu-Zych2013}. 

We note that the selection criteria for LBAs and GPs are not mutually exclusive, therefore certain similarities in their X-ray properties can be expected. 
Namely, Green Peas may form a subset of the LBAs, as they are compact and low mass not only in the UV, but also in the optical light. 
LBAs and GPAs are represented in the BPT diagram of Figure~\ref{fig_bpt}; they are located in a similar region. 
We did not have sufficient data to plot the remaining galaxies of the control sample.

Finally, we added another LBA sample from \citet{Jia2011}, for discussion purposes. Unlike in \citet{Brorby2016}, the LBAs from this sample are not purely star-forming and may contain AGNs -- they fall in the transition region of the BPT diagram 
(the offset from other LBAs and from GPs is well illustrated in Figure~\ref{fig_bpt}).
Therefore, we do not include this sample in our main results of Section~\ref{results}, which are devoted to purely star-forming galaxies. We refer to Jia's sample in Section~\ref{Discussion}, where we explore the possible origins of GP X-ray emission and the role of AGNs. 

\medskip

\subsection{{\rm{XMM-Newton}} Observations and Data Reduction}
We obtained the X-ray data (PI M.\,Ehle) using the 
{\xmm} satellite \citep{2001A&A...365L...1J}. The
three targets were observed in spring 2013.
We report the {\xmm} observational details in Table~\ref{observations}. 
We used both EPIC/PN \citep{pn} and MOS \citep{mos} cameras of {\xmm}, which operated in the Full Frame mode with the thin filter during all of the observations.
The exposure times ranged from 31 to 61\,ks per target. 
However, 
Table~\ref{observations} reports the net exposures after the subtraction of intervals affected by high background flares. The useful exposure of GP3 shrank from 61 to 25.5\,ks, which was insufficient for the detection of the source.

\begin{figure}[tb]
 \includegraphics[width=0.49\textwidth]{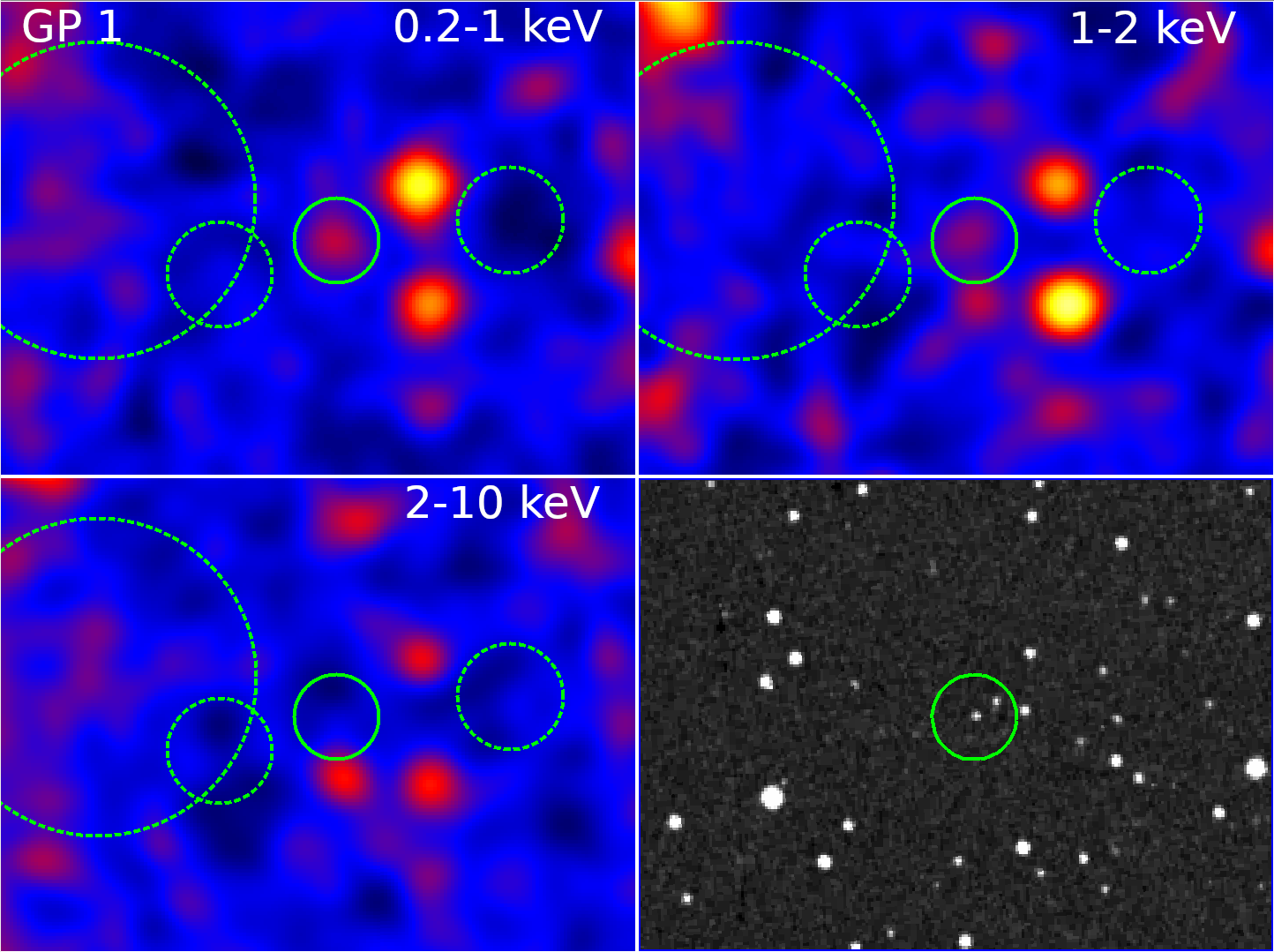} 
 \caption{X-ray images of source regions of GP1 in different energy bands: 0.2--1 keV (top left), 1--2 keV (top right), 2--10 keV (bottom left), and optical SAO-DSS image (bottom right). The size of the images is 5.8 x 4.5 arcmin. The solid circle denotes the extraction region of X-ray photons from the source. The dashed circles are X-ray background extraction regions for PN (smaller circles) and for MOS (large circle). Details of the extracted regions are summarized in the Appendix~\ref{Details_extraction}.}
 \label{image_gp_1}
\end{figure}

We used the Science Analysis System (SAS) software version 15.0 \citep{gabriel2004} to reduce the {\xmm} data. The cleaned event lists were prepared using the SAS commands {\textsc{epchain}} and {\textsc{emchain}}
for the EPIC/PN and MOS detectors, respectively. Only low-background intervals of the rate curve were considered
(i.e. count rate 0.35 cts\,s$^{-1}$ for MOS above 10 keV and 0.4 cts\,s$^{-1}$ in the 10-12 keV energy range for PN).
The source spectra were extracted from circular regions with a radius of 30$\arcsec$ 
around the image coordinates corresponding to the sky position of the sources obtained from the SDSS survey.
For GP1, the extraction region was reduced to 24$\arcsec$ arcsec because of the proximity of another X-ray source that could contaminate the GP1 spectrum (see the X-ray image in Figure~\ref{image_gp_1}).

The X-ray background was measured in nearby source-free circular regions on the same chip of the detector.
For the EPIC/PN detector, we defined the background regions according to the recommendations given 
in the {\xmm} Calibration Technical Note XMM-SOC-CAL-TN-0018 \citep{Smith2016}, so that
the distance from the readout node was similar to that of the source region to ensure comparable low-energy instrumental noise in both regions.
The background region does not contain the same columns as those passing through the source region
to avoid out-of-time events from the source.
For the MOS detectors, there is no such limitation, which allowed us to define
larger background extraction regions than for the PN. 
They were defined as circular regions in the source-free area on the same chip and near the source extraction regions.
Details of the source and background extraction regions are described in the Appendix~\ref{Details_extraction}.

Each EPIC source and background spectra were extracted in the same PI range 0-11999\,eV.
Only patterns less than 4 and 12 were used for the PN and MOS detectors, respectively.
Response and ancillary response matrices were created by the {\textsc{rmfgen}} and {\textsc{arfgen}} tools.
For the PN detector, the response matrices were calculated only up to 12 keV using 2400 energy bins
in order to combine PN, MOS1, and MOS2 spectra together using the {\textsc{epicspeccombine}} tool.
Combined spectra were further analyzed using the XSPEC version 12.9 \citep{Arnaud1996}. 
The $C$-statistics \citep{Cash1976} was used for fitting the (unbinned) data.
A Poisson distribution of the data was assumed as well as a Poisson background.\footnote{The background is accounted for in the analysis using the implementation of the $C$-statistics (also referred to as $W$-statistics), as described at
https://heasarc.gsfc.nasa.gov/xanadu/xspec/manual/XSappendixStatistics.html}


\begin{figure}[tb]
 \includegraphics[width=0.49\textwidth]{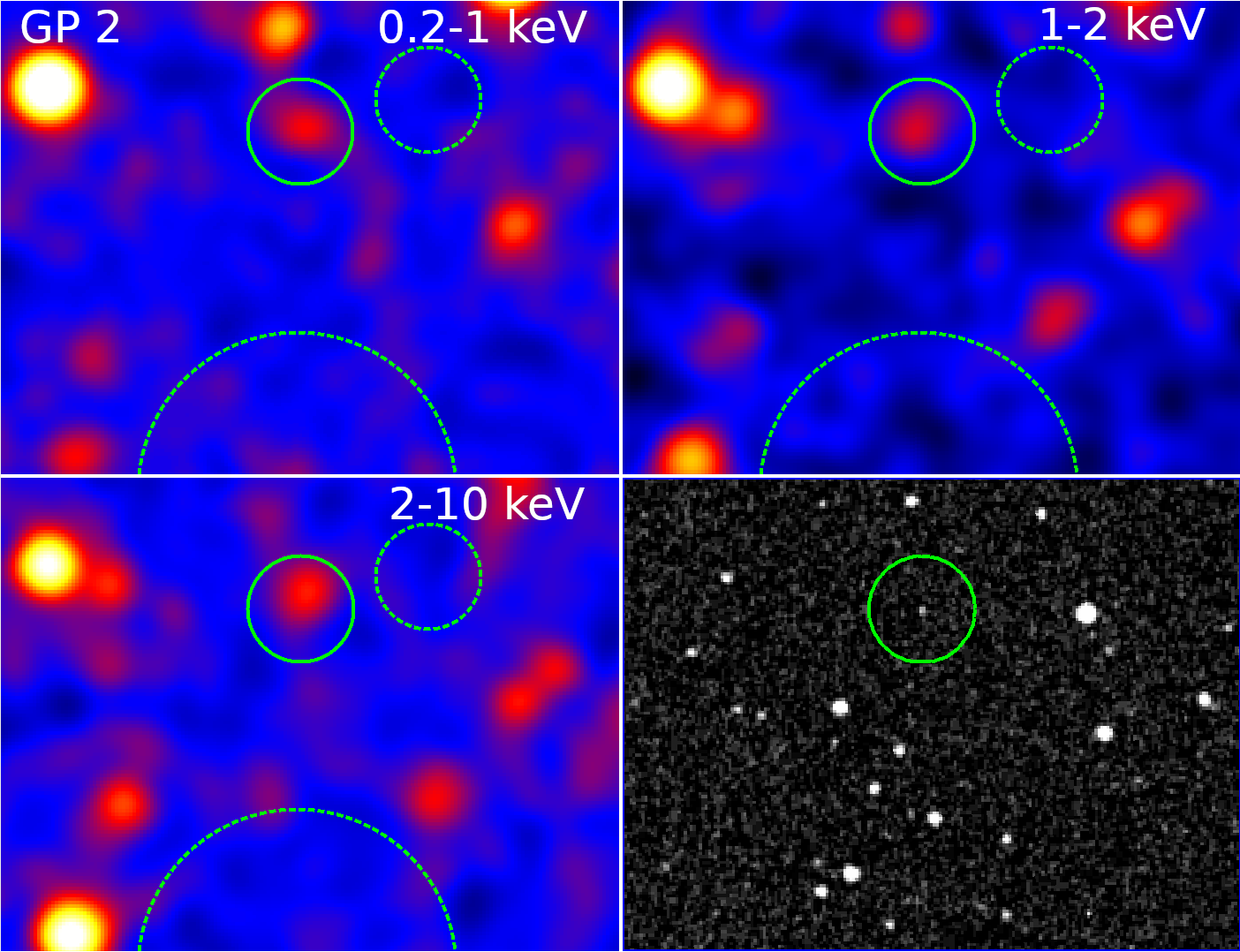}
\caption{Same as in Figure~\ref{image_gp_1} but for GP 2. The source is clearly detected in all energy bands.}
\label{image_gp_2}
\end{figure}

\begin{figure}[tbh]
\includegraphics[width=0.48\textwidth]{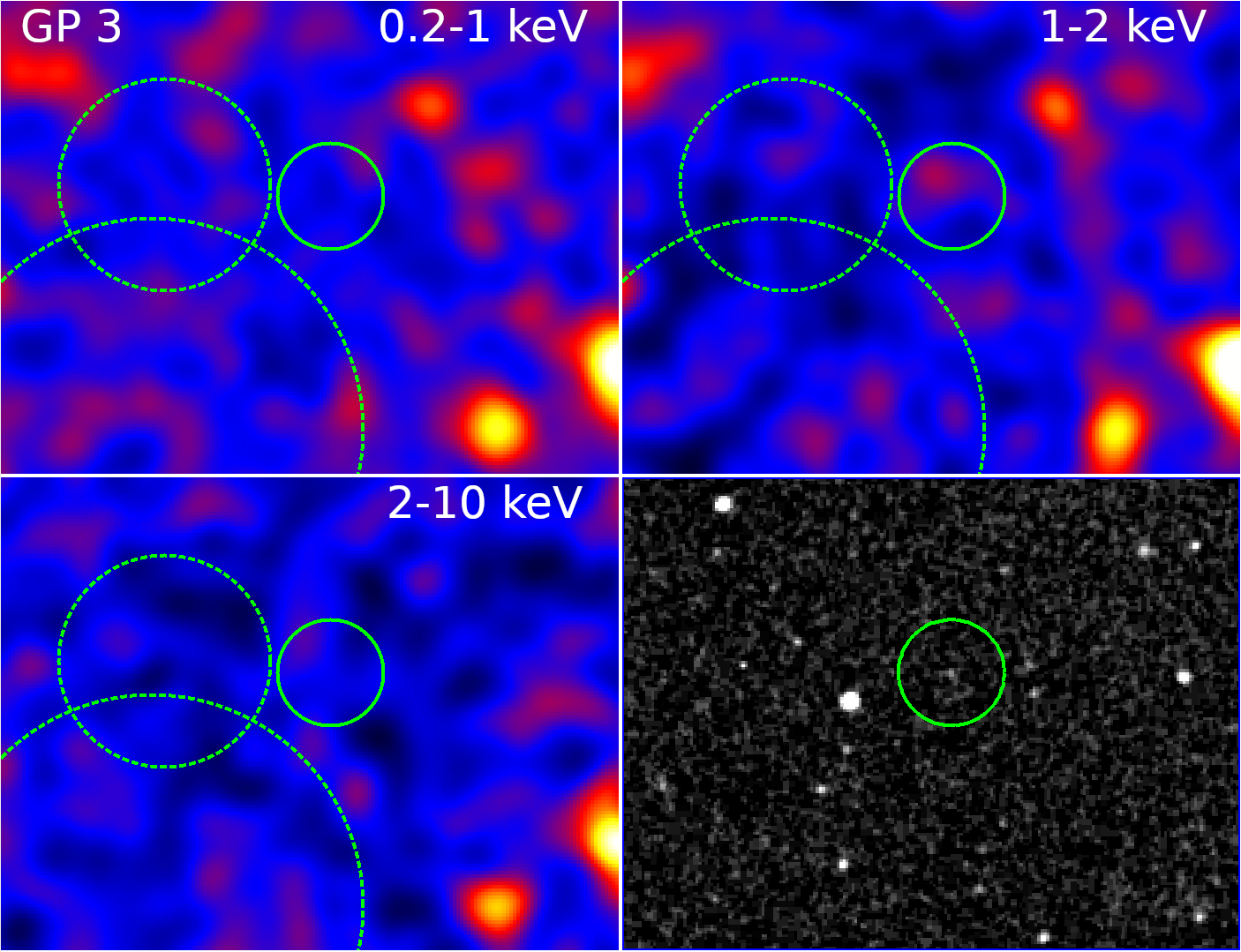}
\caption{Same as in Figure~\ref{image_gp_1} but for GP 3.}
\label{image_gp_3}
\end{figure}

\section{Results}\label{results}

\subsection{X-Ray Detection}

The X-ray images of all source regions are shown in Figures~\ref{image_gp_1}-\ref{image_gp_3} for different energy bands ($0.2-1$\,keV, $1-2$\,keV, $2-10$\,keV and optical SAO-DSS image). 
We followed the SAS data analysis thread "How to generate vignetting-corrected
background-subtracted EPIC images."\footnote{https://www.cosmos.esa.int/web/xmm-newton/sas-thread-images}
For each energy band, an image of the science exposure,
a corresponding scaled out-of-time (OOT) 
image (EPIC-pn only), a scaled Filter Wheel
Closed (FWC) image, and a vignetting-corrected exposure map and mask 
were created. This set of images was used to create the final, i.e. EPIC-pn and MOS combined, background-subtracted and vignetting-corrected science exposure images.
We applied a logarithmic intensity scale to all the X-ray images to allow their direct comparison. GP1 is clearly 
observable as an unresolved point source only in the soft X-ray bands (0.2\,--1\,keV and 1--2\,keV; Figure~\ref{image_gp_1}). 
GP2 is present in all of the energy bands (Fig.~\ref{image_gp_2}).
In contrast, GP3 remained undetected. 
Although there seems to be some increased flux in the 1--2\,keV image of GP3 in Figure~\ref{image_gp_3}, this feature was not found to be statistically significant using the SAS tool  {\textsc{edetect\_chain}}.

Our results were confirmed by the pipeline search detection used for the 3XMM data archive \citep{Rosen2016}. The first two sources were detected and were named as 3XMM J074936.7+333717 (GP1) and 3XMM J082247.4+224145 (GP2).
The significance of the detection expressed by the likelihood is 45 and 63, respectively (see the pipeline manual).
This is well above the recommended threshold of the likelihood parameter (= 10) when a false detection probability is below 0.5\% \citep{Watson2009}, and therefore the detection is confirmed.
GP3 was not detected by the pipeline.

We note that there are two additional sources within the X-ray extraction region in the optical image of GP1. We checked at the NED database\footnote{https://ned.ipac.caltech.edu} that these sources are regular stars in our Galaxy, discovered by the infrared (IR) survey with 2MASS. There is no evidence of the enhancement of the X-ray flux toward the direction to the stars in any X-ray band. Thus, we do not expect any significant contribution of these stars to the detected X-ray emission of GP1.
However, a source of significant X-ray emission occurs close to GP1 at the hard X-ray images (in the bottom direction from GP; see Fig.~\ref{image_gp_1}).  To eliminate any possible contamination of the hard X-ray spectrum of GP1, we reduced the extraction region of the GP1 spectrum to 24 arcsec (instead of 30 arcsec used for the other sources). Because the detection of GP1 above 2\,keV appears to be insignificant, any contamination by that nearby source is ruled out (see more details about the significance of X-ray detection and estimates of possible contamination by background AGNs in Sect.~\ref{xray_significance}).


\begin{table*}[thb!]
\caption{X-Ray flux measurements of Green Pea galaxies.}
\centering
\begin{tabular}{cccccc}
 \rule{0cm}{0.5cm}
 Source & Net Count Rate & \multicolumn{3}{c}{Observed X-ray Flux [$10^{-15}$\,erg\,cm$^{-2}$\,s$^{-1}$]} & X-ray Luminosity [$10^{42}$\,erg\,s$^{-1}$]  \\
 \rule{0cm}{0.5cm}
 &  0.3-10\,keV, [$10^{-3}$ cts/s] & 0.5-2\,keV & 2-10\,keV & 0.5-8\,keV (rest frame) & 0.5-8\,keV (rest frame) \\
 \hline \hline
 \rule[-0.7em]{0pt}{2em} 
SDSSJ074936.7+333716 (GP1) & $3.1 \pm 0.7$ &  $3.1^{+1.0}_{-0.9}$ &  $<1.1$  &  $5.0 \pm 1.5$ & $1.2 \pm 0.4$   \\
 \rule[-0.7em]{0pt}{2em}   
SDSSJ082247.6+224144 (GP2) & $6.4 \pm 0.7$ &   $4.1^{+0.9}_{-0.8}$ &  $4.4^{+3.4}_{-2.2}$ & $8.0^{+1.8}_{-1.5}$ &   $1.2^{+0.2}_{-0.3}$   \\
\end{tabular}
\label{flux} 
 \tablecomments{
 The reported values of 0.5-2 and 2-10\,keV fluxes are observed unabsorbed flux measurements. The flux and luminosity calculated for the 0.5-8\,keV are in the rest frame of the galaxy, i.e. derived from the observed flux in the 0.39-6.28\,keV band for GP 1 and the 0.41-6.58\,keV band for GP2.}
\end{table*}

\begin{figure*}[tbh]
 \includegraphics[width=0.87\textwidth]{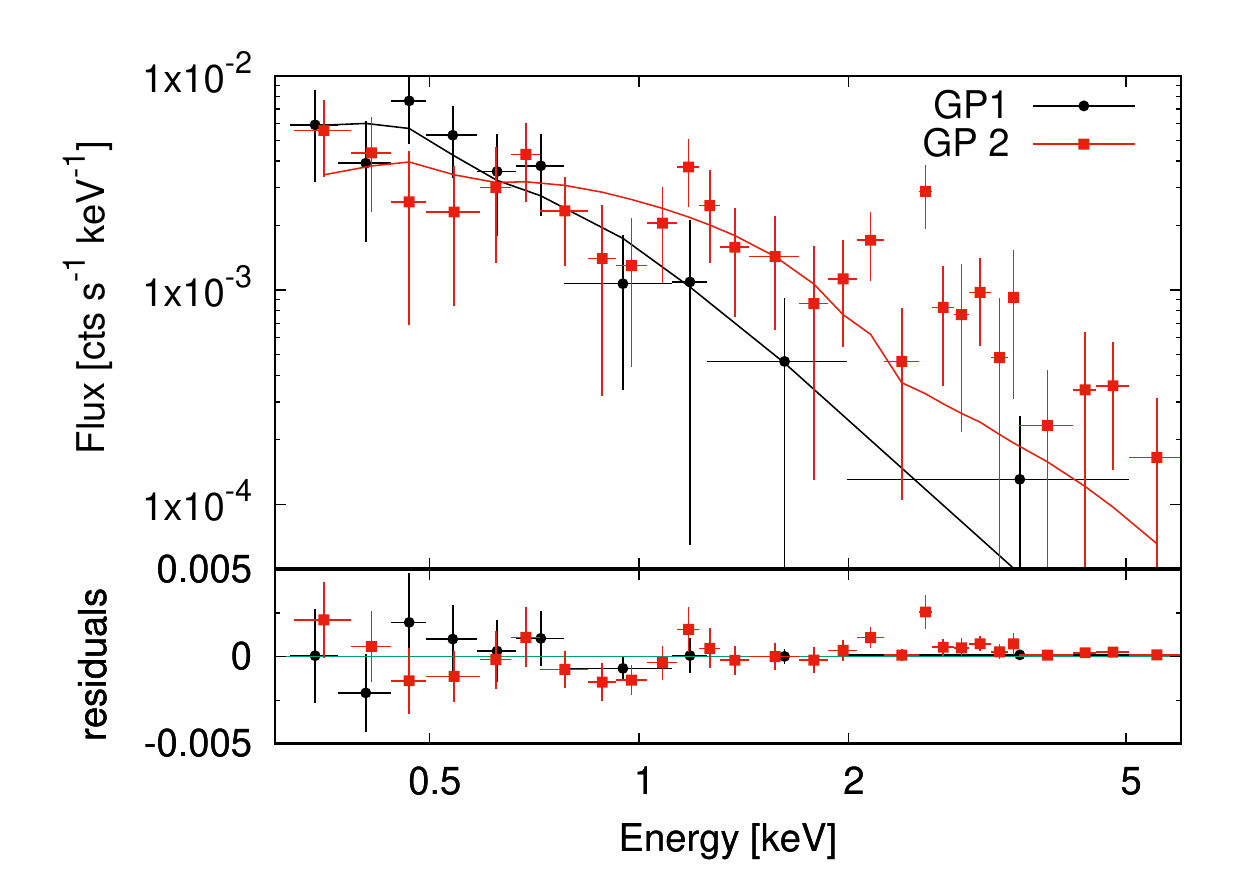}
\centering
\caption{X-ray spectra of GP1 (black circles) and GP2 (red squares), obtained as combined spectra of all {\xmm} EPIC detectors with the best-fit absorbed power-law model overplotted on the data. 
Residuals of the data after the model subtraction (in normalized counts\,s$^{-1}$\,keV$^{-1}$) are shown in the bottom panel. The plotted data are binned so that the bins do not oversample the energy resolution of the instrument and the signal-to-noise ratio is at least 1 in each bin. The binning is used for plotting purposes only; the spectral fitting was done on unbinned data.
}
\label{spectrum_gp}
\end{figure*}

\begin{table*}[tb!]

\caption{Predicted and Measured X-Ray luminosities of Green Pea galaxies.}
\centering
\begin{tabular}{cccccc}
 \hline \hline
 \rule{0cm}{0.5cm}
 Source  &  \multicolumn{4}{c}{$L_{\rm 0.5-8\,keV}$, Predicted from $L_\mathrm{x}$-SFR(-Metallicity) Relations} & $L_{\rm 0.5-8\,keV}$, measured \\
 \rule{0cm}{0.5cm}
&  {\footnotesize{Ranalli et al. (2003)$^{a}$}}  &  {\footnotesize{Mineo et al. (2014)}} & {\footnotesize{Douna et al. (2015)$^{b}$}} & {\footnotesize{Brorby et al. (2016)}}  & {\footnotesize{{\xmm} Observation }} \\

 \rule[-0.7em]{0pt}{2em} 
SDSSJ074936.7+333716 (GP1) & $0.94$ &  $2.35$ &   $1.84$  
& 3.26
&  $12 \pm 4$   \\
\rule[-0.7em]{0pt}{2em}   
SDSSJ082247.6+224144 (GP2) & $0.60$ &  $1.50$ &  $1.17$ 
&  2.51
&   $12^{+2}_{-3}$   \\
 \rule[-0.7em]{0pt}{2em} 
SDSSJ133928.3+151642 (GP3) &  $0.30$ &  $0.75$ &  $0.59$ &  1.29
&   $\le 1.3$ (upper limit)   \\
 \rule[-0.7em]{0pt}{2em} 
 \end{tabular}
\tablecomments{$^{a}$ Ranalli's values were calculated using the $L_\mathrm{x}/{\rm SFR}$ ratio in the original paper and then corrected as in \citet{Mineo2012} to include the whole mass range for SFR.
$^{b}$ Douna's values were calculated using relation $L_\mathrm{x}/{\rm SFR} = 2.09 \times 10^{39}$\,erg\,s$^{-1}$\,yr\,$M_\odot$, which corresponds to the best model at high metallicity, and then corrected (by a factor $3.9/2.6$) following \citet{Mineo2014} to take into account the unresolved emission. The measured luminosity is in intrinsic rest frame of the galaxy.
All luminosities are in units of [$10^{41}$\,erg\,s$^{-1}$].}
\label{luminosity} 
\end{table*}

\subsection{X-Ray Spectral Analysis}

For the first two GPs that are clearly detected, we performed spectral fits in the wide energy range 0.3-10\,keV using XSPEC.
We used a simple absorbed power-law model. The absorption was attributed only to the neutral absorption
in our Galaxy. The column density of the neutral hydrogen was set to values from the Leiden-Argentine-Bonn survey \citep{Kalberla2005},
i.e. $N_\mathrm{H} = 4.65 \times 10^{20}$\,cm$^{-2}$ for GP1 and $N_\mathrm{H} = 4.13 \times 10^{20}$\,cm$^{-2}$ for GP2.
The X-ray continuum of both spectra was well fitted by the power-law model,
resulting in a $C$-statistics 
fit goodness $C=1241$ (1937 degrees of freedom) for GP1
and $C=1116$ (1937 degrees of freedom) for GP2.
The fitting was performed on unbinned data to keep the complete information for the $C$-statistics.  
The X-ray spectra with the best-fit models and the data residuals after the model subtraction are shown in Figure~\ref{spectrum_gp}.
The observational data were rebinned to a signal-to-noise ratio of 1 for a clearer presentation. The data residuals show that the X-ray spectra are well described by the power-law models. The GP2 data points that appear above the model at energies larger than 2\,keV may suggest a higher normalization of the power law and a possible absorption in the soft X-ray band, but the data quality does not allow us to test more complicated models.  
 
The fitted X-ray spectrum of GP1 is significantly steeper than that of GP2.
The photon index of the power-law was found to be $\Gamma = 3.2 \pm 0.7$ 
with the normalization factor $K = (2.7 \pm 0.8) \times 10^{-6}$ for GP1, and $\Gamma = 2.0 \pm 0.4 $ with normalization  $K = (2.8 \pm 0.6) \times 10^{-6}$ for GP2.
This is consistent with the findings on the source detection shown in the image at different energy bands (Figures~\ref{image_gp_1}-\ref{image_gp_3}). GP1 is clearly detected only in the softest energy bands, while GP2 has a significant detection also above 2\,keV. 

The steeper spectrum of GP1 can either be due to the steeper power-law slope, or due to an X-ray excess in the soft X-ray band, or a combination of both. 
We were able to fit the data using the {\sc{apec}} hot plasma model with a resulting $C$-statistic 
fit goodness $C=1248$ (1937 degrees of freedom), only slightly worse than that obtained with the power-law model. The fitted value of the temperature $kT = 0.3^{+0.3}_{-0.1}$\,keV was found to be consistent with typical plasma temperatures in star-forming galaxies \citep[$kT \approx 0.2-0.3$,][]{Mineo2012b}.
However, the measured luminosity of the hot gas would be at least an order of magnitude larger than expected from the empirical prediction
(see Section \ref{discussion_xrb_hotgas} for more details). This is the main reason why we use the power-law model for any further considerations,
while being aware of the data limitation on making any strong conclusions about the X-ray spectral slope of GP1.

Table~\ref{flux} summarizes the X-ray unabsorbed flux measurements in the 0.5--2\,keV 
energy band for GP1 and GP2 (i.e. the targets with a clear detection). 
For GP2, we also report the flux in the 2--10\,keV energy band.
For GP1, whose X-ray spectrum was found to be significantly steeper, we only 
constrained an upper limit in the  2--10\,keV energy band.

We then derived the unabsorbed flux in the galaxy rest frame 0.5--8\,keV energy band for each target, which will allow a comparison with other star-forming galaxies. 
This translates into the 0.39--6.28\,keV energy band for GP 1 and the 0.41--6.58\,keV band for GP2.
The X-ray luminosity was then constrained as $L = 4\pi D_{\rm L}^2\,F_{\rm unabsorbed}$
using the luminosity distance calculated from the cosmological redshift.
The luminosity distance is $D_{\rm L} = 1437.6$\,Mpc for GP1 and $D_{\rm L} = 1103.1$\,Mpc for GP2 (using the NED cosmological calculator). 

We note that the intrinsic 0.5--8\,keV X-ray flux of GP1 was extrapolated from the soft ($<\!2$\,keV) X-ray band and its value depends on the employed model. Namely, for the {\sc apec} model, the extrapolated flux would be a factor of 1.5 lower
than the flux determined from the power-law model. 
Nevertheless, it is highly unlikely that  
the power-law component would be completely missing in the galaxy spectrum and, in fact, 
the hard-band X-ray flux could be even larger if we assumed the standard photon index for star-forming galaxies ($\Gamma \approx 1.9-2$). Therefore, we expect that the flux extrapolated from the steep power-law model represents a realistic and rather conservative estimate of the flux.

Finally, we performed the analysis of the shortened observation of GP3. 
We only constrained an upper limit for the X-ray flux based on the non-detection of the source.
We derived the flux upper limit $F_{\rm 2-10\,keV} = 7.2 \times 10^{-16}$\,erg\,cm$^{-2}$\,s$^{-1}$,
corresponding to $L_{\rm GP3,\,2-10\,keV} = 0.77\times10^{41}$\,erg\,s$^{-1}$
(assuming a power-law spectrum with $\Gamma = 2$).

\subsection{Ancillary Optical Spectra Analysis}
\label{optical_analysis}

Interpretation of the X-ray fluxes measured in star-forming galaxies 
relies on multi-wavelength data, namely on independent SFR, mass, 
and metallicity estimators in the optical, UV, and/or infrared wavelengths.
For GPs, various estimates were published by 
\citet{Cardamone2009, Amorin2010, Izotov2011}, and in 
online catalogs such as MPA-JHU \citep{Brinchmann2004}.
The results available for GP1, GP2 and GP3 vary depending on the method applied in each study. In this paper, we adopted the results and methods that were most compatible with those used in our control sample in order to allow for a direct comparison between the galaxies. Whenever possible, we adopted the corresponding values from the literature. Where necessary, we re-derived the parameters
using the SDSS Data Release~13 spectra \citep[DR13;][]{Albareti17}\footnote{https://dr13.sdss.org/optical/spectrum}. The SDSS optical fibers of $3\arcsec$ in diameter 
encompass most of the signal from each GP due to their compactness.
We measured the continuum-subtracted emission-line fluxes using double-Gaussian fits to the line profiles.

The literature does not provide consistent results on the GP metallicity, measured in terms of the oxygen abundance 12\,+\,log\,(O/H). 
This is a well-known effect that is due to the different calibrations of the metallicity estimators. \citet{Kewley2008} derived conversions between them, which are, however, valid in the statistical sense and may not be precise for individual objects. We have therefore recomputed 12\,+\,log\,(O/H) directly from the SDSS spectra, using a method compatible with that used in our control sample. We applied the O3N2 calibration derived by 
\citet{Pettini2004} based on the 
[O\,{\sc iii}]$\lambda$5007/H$\beta$ and [N\,{\sc ii}]$\lambda$6583/H$\alpha$ emission line ratios.
We obtained metallicities (Table~\ref{SFR_table})
that are consistent with those derived by \citet{Izotov2011} using the direct temperature method, and are lower than the [O\,{\sc ii}]- and [N\,{\sc ii}]-based measurements of \citet{Cardamone2009}. See Section~\ref{discussion-metallicity} for more details.

We adopted the GP SFRs (Table~\ref{SFR_table}) from \citet{Cardamone2009}, 
who derived them from the H$\alpha$ line flux \citep{Kennicutt1998}, 
corrected for the underlying stellar absorption and interstellar extinction.
Even though the SFRs of the control sample were measured using combined near-UV and infrared (NUV+IR) fluxes, we show by testing different empirical conversions between H$\alpha$ and NUV+IR in Section~\ref{Discussion} that the choice of the H$\alpha$ SFR does not  significantly affect our results. Therefore, for simplicity, we present our main results using the H$\alpha$ data.

Finally, we adopted the GP stellar masses from the MPA-JHU galaxy catalog\footnote{http://wwwmpa.mpa-garching.mpg.de/SDSS/DR7/} \citep{Brinchmann2004}, obtained by the SDSS optical spectral energy density (SED) fitting (Table~\ref{SFR_table}). We used the total mass (as opposed to the fibre mass).
This choice is consistent with that made by \citet{Brorby2016} for their set of LBAs. 
These LBAs are part of our comparison sample and many of their properties are similar 
to those of the GPs, hence a consistent choice of methods is most critical here.  
In addition, we checked other independent mass estimates for the GPs \citep{Cardamone2009,Izotov2011}, and obtained consistent results within a factor of two for GP1, and within a factor of five for GP2 and GP3 (in linear scale). All of these derivations were based on the optical SED fitting, while no other mass estimators were available: the GPs were not detected in the 2MASS survey due to the faintness of their old stellar populations. Nevertheless, we only use the stellar masses here for deriving the specific star formation rates (sSFR, i.e. SFR per unit stellar mass), and we have checked that the uncertainty in mass does not introduce any effects in our results and interpretation. 
The sSFR is useful for determining whether the X-ray luminosity is dominated by the population of high-mass or low-mass X-ray binaries \citep{Mineo2012}. The sSFR of our GPs 
(Table~\ref{SFR_table}) are all significantly larger than the threshold log(sSFR)$ = -10$\,[yr$^{-1}$] \citep{Mineo2012}, indicating a very high SFR per stellar mass and thus suggesting a clear dominance of the 
high-mass X-ray binaries in the X-ray flux. 
The division between the HMXB-dominated and LMXB-dominated galaxies is only valid if no active galactic nucleus is present. We have checked the classification for our sample, based on the optical emission-line ratios and using the BPT diagram \citep{Baldwin1981}, and we found no indication of an AGN (but see Section~\ref{sec_AGN} for further details).

\begin{figure*}[tb!]
 \centering
\includegraphics[width=0.9\textwidth]{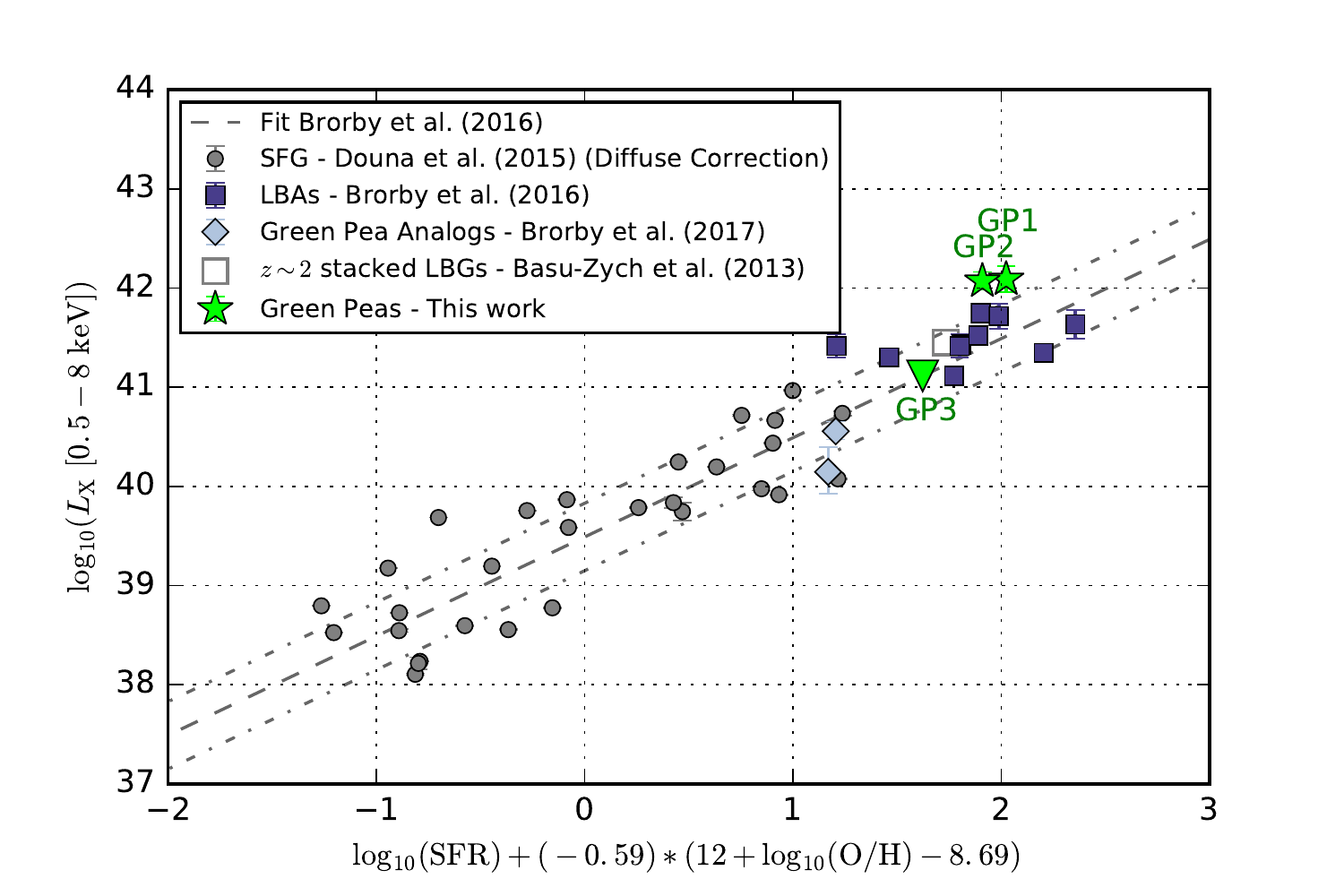}
\caption{
GPs of our sample overplotted on the \citet{Brorby2016} diagram
of the L$_{\rm x}$--SFR--metallicity plane. GP1 and GP2 are shown by green-filled stars. The upper limit constrained for GP3 is shown by a green triangle. The light-blue diamonds correspond to the Green Pea analogs \citep{Brorby2017}, 
the dark-blue squares are the Lyman-break analogs (LBAs) \citep{Brorby2016} and 
the gray circles correspond to the star-forming galaxies (SFGs) from 
\citep{Douna2015},
corrected for the diffuse X-ray emission. The $z=1.9-2$ stacked Lyman-break galaxies (LBGs) from \citet{Basu-Zych2013a} have also been included for comparison, with metallicity adopted from \citet{Basu-Zych2013}. The dashed and dashed-dotted lines represent the best-fitting linear trend and its $1\sigma$ 
deviation, respectively, derived by \citet{Brorby2016}. The X-ray luminosity is in units of erg s$^{-1}$, and the SFR in units of $ M_{\odot} {\rm yr}^{-1}$.
}
\label{brorby}
\end{figure*}

\begin{figure*}[tb!]
\begin{tabular}{ll}
\includegraphics[width=0.485\textwidth]{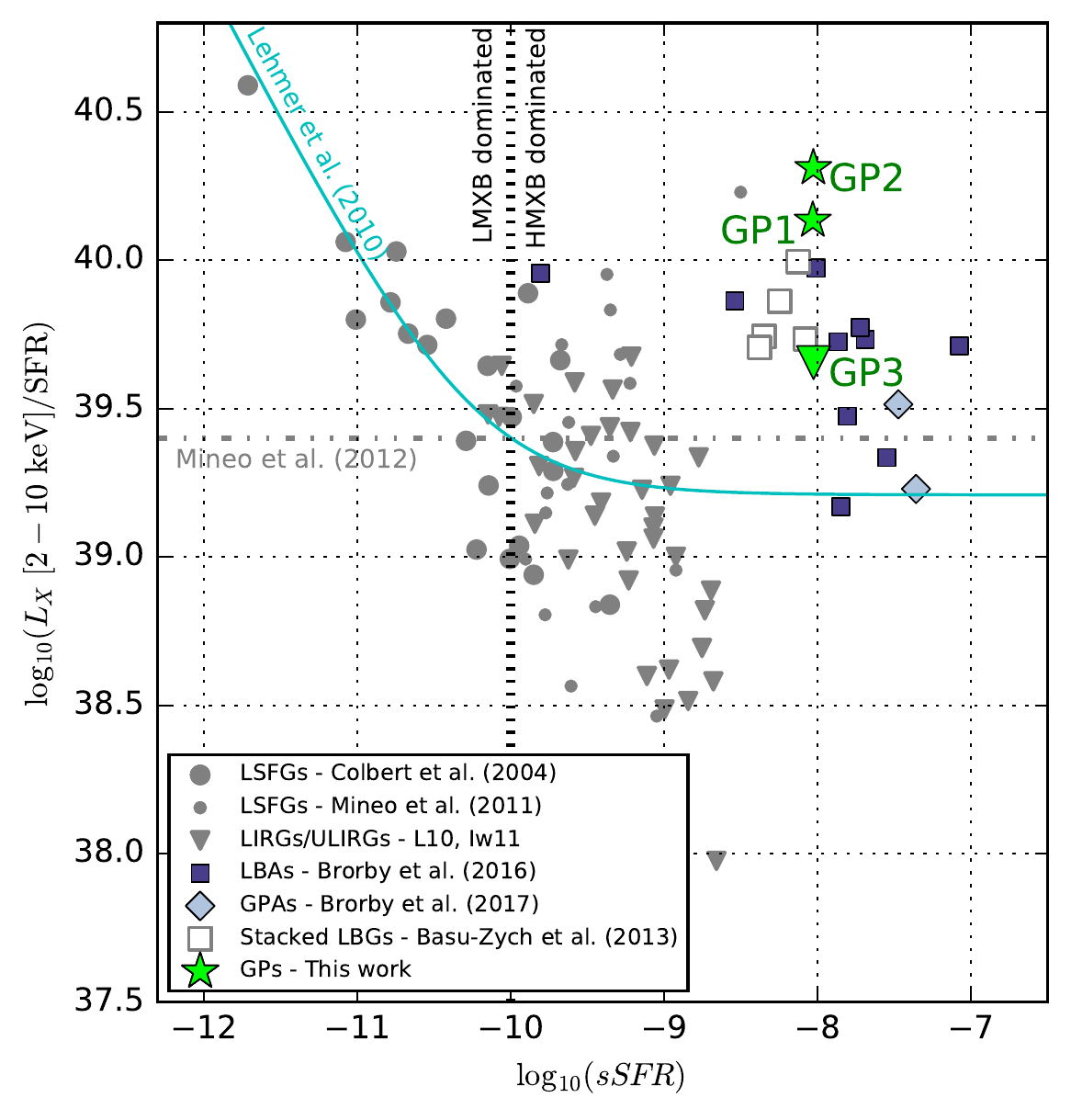}
&
\includegraphics[width=0.51\textwidth]{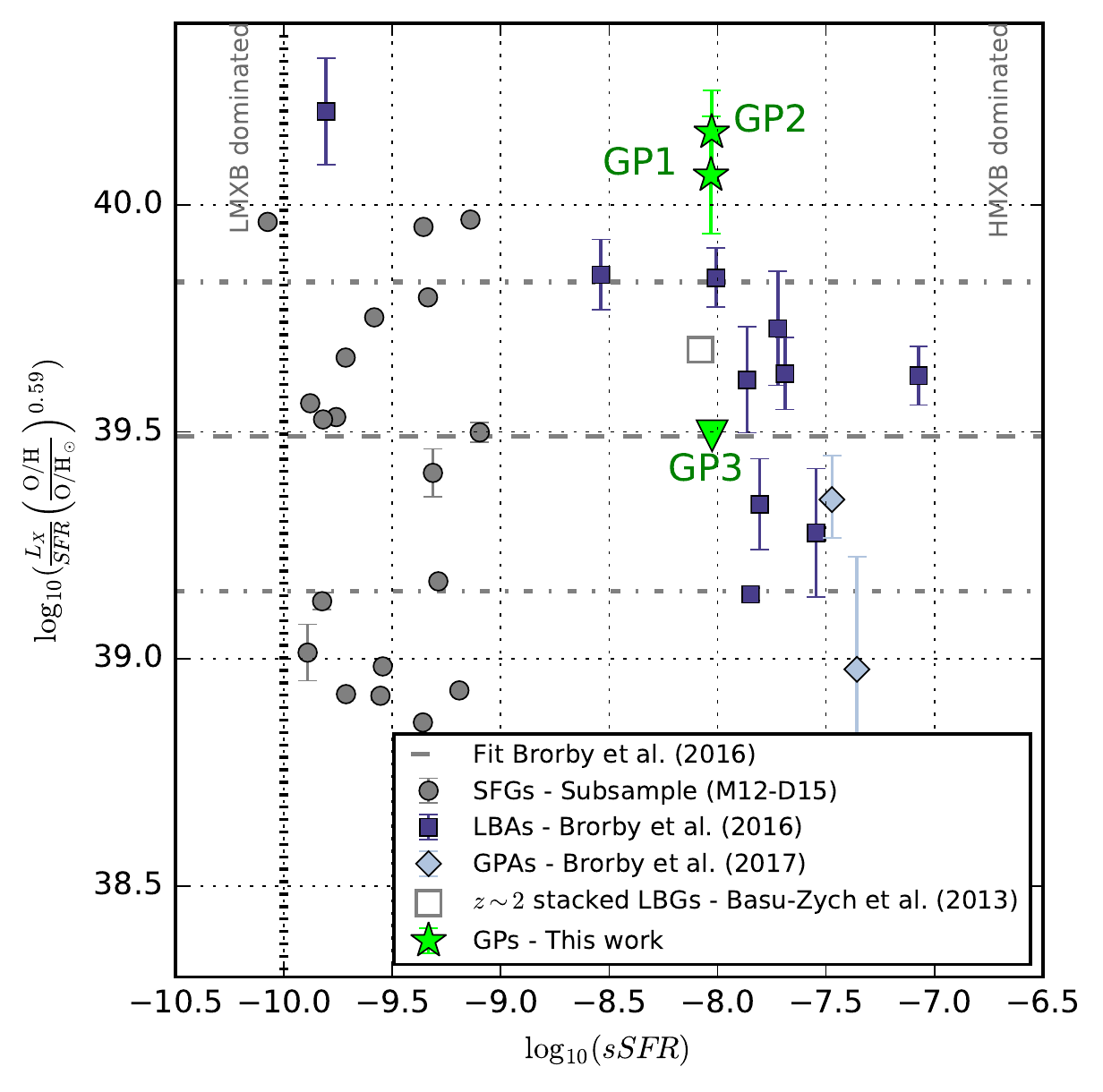}
\end{tabular}
\caption{{\textbf{Left:}} Green Pea galaxies overplotted in the diagram
of X-ray luminosity per unit SFR, as a function of specific SFR, 
originally proposed by \citet{Basu-Zych2013}. 
The stacked LBGs (different redshifts, open squares) are from \citet{Basu-Zych2013a}; Luminous Star-Forming Galaxies (LSFGs, grey circles) are from \citet{Colbert2004} and \citet{Mineo2012}; LIRGs and ULIRGs (grey triangles) are from \citet{Lehmer2010} and \citet{Iwasawa2011}.
We added LBAs (dark-blue squares) from \citet{Brorby2016} and Green Pea Analogs from \citet{Brorby2017}.
{\textbf{Right:}} 
X-ray luminosity (0.5-8 keV) with respect to the SFR scaled by the metallicity according to the correlation present in \citet{Brorby2016}. The gray points correspond to the star-forming galaxies in \citet{Mineo2012} for which the metallicities were available in \citet{Douna2015}, while the blue squares are the LBAs from \citet{Brorby2016} and the diamonds represent GPAs from \citet{Brorby2017}. 
The gray open square corresponds to the $z = 1.9-2$ stacked LBGs from \citet{Basu-Zych2013a}, with metallicity adopted from \citet{Basu-Zych2013}. GP1 and GP2, which have a reliably measured X-ray flux are above the correlation, while GP3 with its upper limit estimate lies exactly on the correlation. The X-ray luminosity is in units of erg s$^{-1}$, the SFR in units of $M_{\odot} {\rm yr^{-1}}$ and the sSFR is in yr$^{-1}$.}

\label{basuzych}
\end{figure*}

\subsection{Elevated X-Ray Luminosity in GPs}
\label{elevated}

We here compare the  X-ray luminosities that we measured for the three GPs with their 
expected values, derived
from relations reported in the literature. 
\citet{Ranalli2003} provided relations
between the SFRs of a sample of nearby star-forming galaxies and their total X-ray luminosity in the soft and hard X-rays, $L_{\rm x, 0.5-2\,keV} \, [{\rm erg \ s^{-1}}] \approx 4.55 \times 10^{39} \, {\rm SFR \ [M_{\odot} \ yr^{-1}]}$, and $L_{\rm x, 2-10\,keV} \, [{\rm erg \ s^{-1}}] \approx 5 \times 10^{39}\, {\rm SFR \ [M_{\odot} \ yr^{-1}]},$ respectively. However, the SFR here refers to stars more massive than $5M_{\odot}$.
\citet{Mineo2012} thus computed an extension to lower masses, which yields 
SFR\,$(0.1\!-\!1000\,M_{\odot}) = 5.5\,\times $\,SFR\,$(>\!\!5M_{\odot})$,
using the Salpeter initial mass function (IMF).  
This leads to the L$_x$-SFR relation 
$L_{\rm x, 0.5-8\,keV} \, [{\rm erg \ s^{-1}}] \approx 1.6 \times 10^{39}\, {\rm SFR \ [M_{\odot} \ yr^{-1}]}$.
A slightly larger X-ray luminosity was constrained for the emission of the population of X-ray binaries from the analysis by \citet{Grimm2003}, $L_{\rm x, 0.5-8\,keV}^{\rm XRB} \, [{\rm erg \ s^{-1}}] \approx 2.8 \times 10^{39} \, {\rm SFR \ [M_{\odot} \ yr^{-1}]}.$ 
\citet{Mineo2012} rescaled their relation to consider updated calibrations of the IR and radio SFR estimators and obtained 
$L_{\rm x, 0.5-8\,keV}^{\rm XRB} \, [{\rm erg \ s^{-1}}] \approx 2.6 \times 10^{39}\, {\rm SFR \ [M_{\odot} \ yr^{-1}]}.$ 
\citet{Gilfanov2014} derived a similar relation, 
$L_{\rm x, 0.5-8\,keV} \, [{\rm erg \ s^{-1}}] \approx 2.5 \times 10^{39}\, {\rm SFR \ [M_{\odot} \ yr^{-1}]}$. 
%
\citet{Mineo2014} then studied the total X-ray emission of unresolved star-forming galaxies (unlike the previous works that considered resolved sources), 
and obtained
%
a relation $L_{\rm x, 0.5-8\,keV} \, [{\rm erg \ s^{-1}}] 
\approx 4 \times 10^{39} \, {\rm SFR \ [M_{\odot} \ yr^{-1}]}$. 
By comparison with the resolved X-ray sources of \citet{Mineo2012}, 
they inferred that the hot gas emission represented approximately 
one third of the total $L_\mathrm{x}$.
The GPs studied in the present paper are unresolved due to their relatively large distances, and therefore any appropriate comparison with other samples should also consider the unresolved emission. 

The effects of  metallicity, and thus more complex $L_{\rm x}$--SFR--metallicity relations, were subsequently considered by \citet{Fragos2013a, Basu-Zych2013, Brorby2014, Douna2015} and \citet{Brorby2016}. 
By means of Bayesian inference, \citet{Douna2015} fitted Monte Carlo models to clarify the dependence  of the size and luminosity of HMXB populations on metallicity in nearby star-forming galaxies. They showed that the number of HMXBs is an order of magnitude higher in galaxies with 12+log(O/H)\,$\lesssim$\,8 than in near-solar metallicity galaxies. Moreover, \citet{Brorby2016} found an enhancement of the total X-ray luminosity in their LBA sample with respect to solar metallicity galaxies and proposed a possible $L_\mathrm{x}$--SFR--metallicity plane. Their best fit to the data, which agrees with the results of \citet{Douna2015} corrected for the unresolved emission, corresponds to $ {\rm log} (L_{\rm x})={\rm log \ (SFR)} + (-0.59\pm0.13)\ {\rm log \ ((O/H)/(O/H)_\odot)}+(39.49\pm0.09)$ with a dispersion of 0.34 dex.

The different theoretical predictions from L$_{\rm x}$-SFR(-metallicity) relations for the GPs in this work are summarized in Table~\ref{luminosity}.
The predicted values of $L_{\rm 0.5-8\,keV}$ for GP1 are in the range $0.94-3.26 \times 10^{41}$\,erg\,s$^{-1}$,
while the measured value is about an order of magnitude larger,  $L_{\rm 0.5-8\,keV} \approx  1.2 \times 10^{42}$ \,erg\,s$^{-1}$. 
A similar result was obtained for GP2, the predicted value was $L_{\rm 0.5-8\,keV} \approx 0.6-2.51 \times 10^{41}$\,erg\,s$^{-1}$,
while the {\xmm} measured value is $L_{\rm 0.5-8\,keV} \approx 1.2 \times 10^{42}$\,erg\,s$^{-1}.$
The excess in both sources is thus of the order of $\approx\!10^{42}$\,erg\,s$^{-1}$.
For GP3, only a luminosity upper limit could be established, and 
is consistent with the L$_{\rm x}$--SFR--metallicity relation by \citet{Brorby2016}.


\subsection{GPs in the $L_\mathrm{x}$-SFR-Metallicity plane}
\label{lx--SFR--met}

The X-ray luminosity excess in both detected GPs is clearly visible in 
Figure~\ref{brorby}. The plot is based on the diagram originally proposed by \citet{Brorby2016}, where we added the points corresponding to the GP galaxies of our sample. We show the relation that \citet{Brorby2016} derived for the 
$L_\mathrm{x}$--SFR--metallicity plane (see Section~\ref{elevated}) and we include the control samples of \citet{Douna2015,Brorby2016} and \citet{Brorby2017}, described in 
Section~\ref{sample}. \citet{Douna2015} reported X-ray luminosities of resolved X-ray sources for their sample. Therefore, we applied a correction to account for the diffuse X-ray emission in these galaxies to be compatible with the rest of the sample, where the sources are unresolved. 
We multiplied the sum of their X-ray emission by a factor of $4/2.6,$ following the $L_\mathrm{x}$ vs. SFR relation of \citet{Mineo2012} and \citet{Mineo2014}. An analogous correction was applied by \citet{Brorby2016} and \citet{Brorby2017}, and we thus adopted their values for the LBAs and the GP analogs.  
In addition, GPs are considered to be analogs of $z\!\gtrsim\!1$ galaxies, and 
therefore we add the stacked $z\!\sim\!2$ LBGs of \citet{Basu-Zych2013a} for comparison. 
We adopted their mean rest-frame 2-10\,keV X-ray luminosity  
and their rest-frame UV SFR 
from \citet{Basu-Zych2013a}, and their oxygen abundance from \citet{Basu-Zych2013}.
We converted the X-ray luminosity to the 0.5-8\,keV band, assuming $\Gamma\!=\!1.9,$ in order to be compatible with the other galaxy samples.
We also performed the correction for diffuse emission.
\citet{Basu-Zych2013a,Basu-Zych2013} did not provide metallicities for LBGs at other redshifts, which hence could not be represented in the plot.

GPs occupy a similar region of the ``Brorby plot'' as LBAs and LBGs in terms of metallicity and SFR; however, in the X-ray luminosity, GP1 and GP2 are superior to any other group of the control sample.
Both of our X-ray-detected GPs are above the \citet{Brorby2016} line, which indicates an X-ray excess with respect to star-forming galaxies that were used for deriving the relation. High-$z$ LBGs, which we have added to the original plot, comply with the relation within $1\sigma.$ 
In contrast, the galaxies that were selected as Green Pea analogs by \citet{Brorby2017} are shifted to much lower $L_\mathrm{x},$ below the \citet{Brorby2016} line. 
GP1 and GP2 deviate from the trend by more than $1\sigma.$
Despite the fact that the deviation does not reach $3\sigma,$ it is definitely
worth exploring since the X-ray excess is of the order of $10^{42}$\,erg\,s$^{-1}$, suggesting extremely energetic processes. This excess, if due to star formation, would correspond to an X-ray-derived SFR\,$\gtrsim\!300$\,$M_\odot$\,yr$^{-1}$ in GP1 (and analogously for GP2) according to \citet{Brorby2016} relation, i.e. a factor of six larger than measured (cf. Table~\ref{SFR_table}). This would translate, according to the relation for the HMXB number $N_{\rm HMXB}\!\approx\!13 \times$\,SFR \citep{Gilfanov2014}, to an excess of about 3000 ($13*\Delta SFR = 13*250 \approx 3000$) high-mass X-ray binaries per galaxy, which can hardly be attributed to a scatter in a standard X-ray binary population. We will discuss possible explanations of the observed X-ray excess in Section~\ref{Discussion}. 
On the other hand, the GP3 upper limit is consistent with the relation found by 
\citet{Brorby2016}, and therefore this property is not universal for all GPs.

GPs are low-mass, compact galaxies with a large SFR.  
As the SFR scales with mass, it is not immediately obvious from the 
\citet{Brorby2016} plane if the galaxy type plays a role in the GP X-ray excess. 
A useful plot was proposed by 
\citet{Basu-Zych2013}, which 
shows the X-ray luminosity per 
unit SFR as a function of the 
%
specific star formation rate. 
High sSFR is characteristic for galaxies with a prevalence of young stellar populations, hence  HMXBs. In contrast, sSFR smaller than $10^{-10}$\,$M_\odot$\,yr$^{-1}$ 
is a signature of populations older than a billion years dominated by LMXBs.
If SFR were the only parameter determining the X-ray luminosity, 
the $L_\mathrm{x}$/SFR ratio would be constant for HMXBs.
We adopted the \citet{Basu-Zych2013} diagram and also their data that comprise star-forming galaxies from \citet{Colbert2004} and \citet{Mineo2012}, 
(ultra-)luminous infrared galaxies (LIRGs, ULIRGs) from \citet{Lehmer2010} and \citet{Iwasawa2011}, stacked LBGs from \citet{Basu-Zych2013a}, and  theoretical/empirical curves from \citet{Lehmer2010} and \citet{Mineo2012}. We added the recent sample of LBAs from \citet{Brorby2016} that superseded the LBAs of \citet{Basu-Zych2013}, the sample of GPAs from \citet{Brorby2017}, and our sample of GPs. 
We converted all the X-ray luminosities
to the 2\,--\,10 keV band following \citet{Brorby2016}: we applied the factor 0.654 that corresponds to the slope $\Gamma = 1.9$ and the hydrogen column density $N_\mathrm{H} = 3 \times 10^{20} {\rm cm}^{-2}.$ We extracted the masses for the GP analogs from the MPA-JHU database, 
consistently with the GP and LBA samples. We present the diagram 
in the left panel of Figure~\ref{basuzych}.

\citet{Basu-Zych2013} noted that LBAs have an excess of $L_\mathrm{x}$ per unit SFR as compared to other nearby star-forming galaxies; they lie well above the empirical predictions of \citet{Lehmer2010} and \citet{Mineo2012},
and above other galaxies with a similar sSFR, including the ULIRGs.  
In this respect, LBAs were found to be more similar to the distant LBGs than to the local star-forming galaxies. The left panel of our Figure~\ref{basuzych} shows 
that the $L_\mathrm{x}$/SFR excess is yet more pronounced for 
GP1 and GP2. This fact confirms that some GPs are extreme versions of the LBAs.
In contrast, the GP analogs from \citet{Brorby2017} do not show deviant 
$L_\mathrm{X}$ per unit SFR as compared to other samples.
Their sSFR is among the largest in this plot ($\sim\!10^{-7.5}$\,yr$^{-1}$), while their $L_\mathrm{x}$/SFR is one order of magnitude lower than in GP1 and GP2. 
We thus conclude 
that sSFR alone is not a sufficient parameter determining the X-ray emission.  

The original \citet{Basu-Zych2013} diagram did not include
any metallicity dependence. Therefore, 
in the next step, we modified their plot while taking into account the 
metallicity-based results of \citet{Brorby2016}. This resulted in a multiplicative factor
$\left(\frac{O/H}{O/H_\odot}\right)^{0.59}$ applied to the $L_\mathrm{x, 0.5 - 8 keV}$/SFR ratio. This quantity should be constant according to the correlation of \citet{Brorby2016}. 
Note that we used the X-ray luminosity in the 0.5--8\,keV band rather than its extrapolation to the 2--10\,keV band because most of the sources are very soft in X-rays and any extrapolation to the hard band may introduce systematic uncertainties.
We present the new plot in the right panel of Figure~\ref{basuzych}. The horizontal lines correspond to the \citet{Brorby2016} relation and its $1\sigma$ deviation as in Figure~\ref{brorby}. 
We show the position of our GPs 
in comparison to the LBAs from \citet{Brorby2016}, 
{GPAs from \citet{Brorby2017}, the 
stacked $z\!\sim\!2$ LBGs from \citet{Basu-Zych2013a}}
as in the previous plot, and star-formation galaxies from 
\citet{Mineo2012}, for which stellar masses were provided by the authors.
The control samples are hence slightly 
different between the two panels of this Figure. 
Most of the control sample galaxies, including the LBAs and LBGs, lie within the $1\sigma$ uncertainty of the Brorby et al. correlation, which indicates that the LBA X-ray excess reported by \citet{Basu-Zych2013} can be explained by the metallicity. 
The stacked $z\sim2$ LBGs have sSFR similar to most of the studied LBAs; and they are located slightly above the Brorby sequence, similar to most of the LBAs.
In contrast, GP1 and GP2 are clearly above the $1\sigma$ limit, which leaves their interpretation less straightforward. 
Similar to the previous plots, GP1 and GP2 outperform the other galaxy samples in this diagram.

Some galaxies of the control sample reach beyond the $1\sigma$ limit, both above and 
below the Brorby et al. line. Namely, the GP analogs have a similar or larger sSFR than GPs, but lie below the Brorby relation. 
Other control-sample galaxies occupy an interval of lower sSFR than the GPs and are closer to the LMXB limit.  

%
We note that the only LBA situated above the \citet{Brorby2016} correlation by more than 1$\sigma$ is KUG 0842+527, also known as the radio galaxy FIRST J084602.3+523158. The X-ray overabundance could be explained if the radio emission was coming from a relativistic jet launched from the central 
super-massive black hole region. However, this galaxy was classified as a starburst galaxy by \citet{Best2012}, not as an AGN. Another explanation of the enhanced X-ray flux could thus be the contribution of LMXBs. This galaxy has a total stellar mass $M_*\!\sim\!10^{11}\,M_\odot$, i.e.  
two orders larger than the GPs, and its position in the diagram is close to the border separating the LMXB- and HMXB-dominated sources. According to \citet{Gilfanov2014}, $L_{\rm x, LMXB} \approx 10^{39} \frac{\rm{M}_*}{10^{10}\rm{M}_\odot}$, corresponding to $L_{\rm x, LMXB} \approx 10^{40}$\,erg\,s$^{-1}$ for KUG 0842+527.
Nevertheless, this LMXB contribution represents only $\sim\!20\%$ of the $L_\mathrm{x}$ predicted for the HMXBs in KUG 0842+527, given its   
SFR = 18.8\,$M_\odot$\,{yr}$^{-1}$ \citep{Brorby2016}. In any case, the reasons for the location of this LBA galaxy above the empirical sequence are probably different from those determining the position of GP1 and GP2.


\begin{figure}[tb!]
\includegraphics[width=0.5\textwidth]{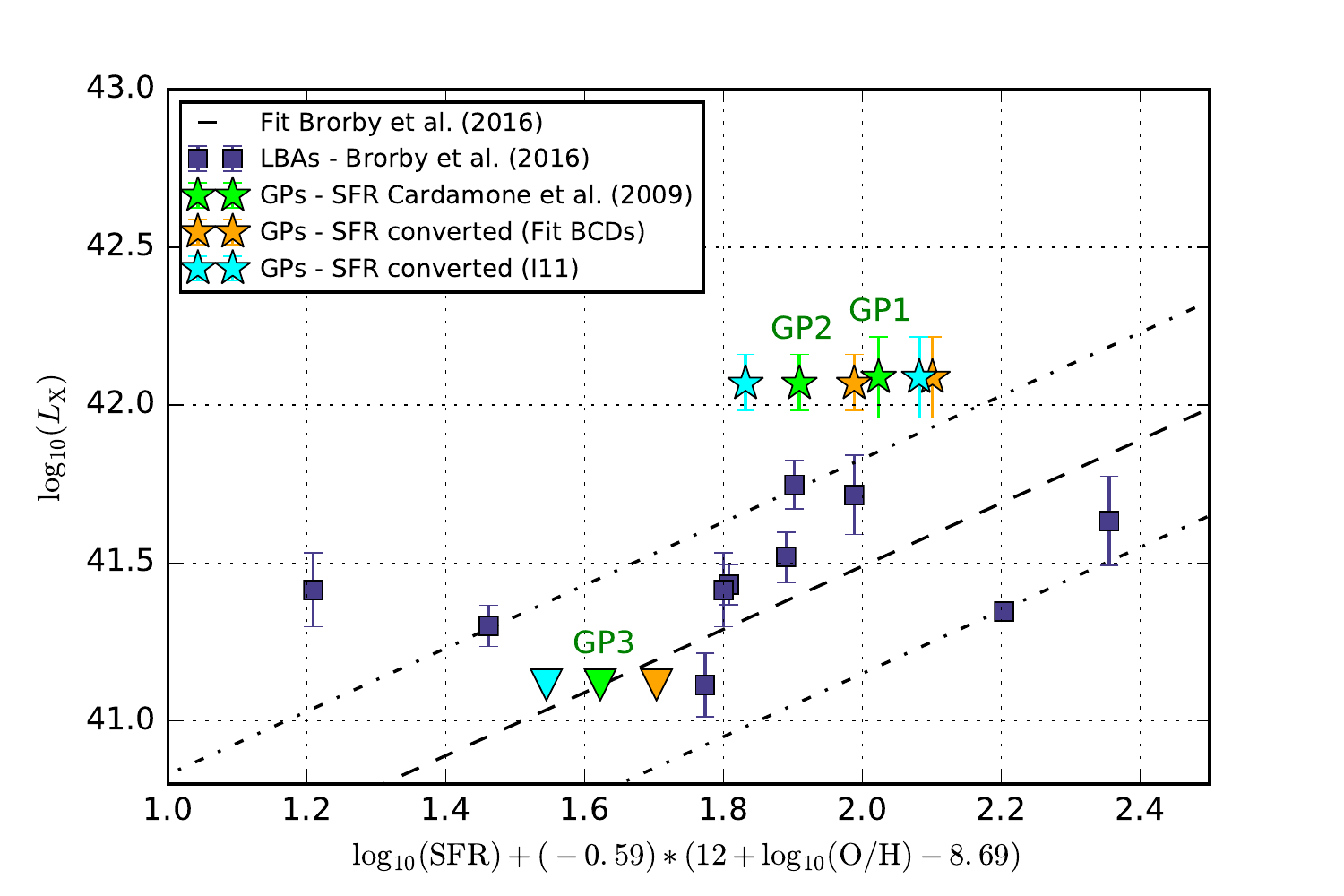}
\caption{Detailed look at the plot presented in Figure~\ref{brorby} showing GPs with different SFR estimates (see the main text for the details).}
\label{SFR_comparison}
\end{figure}

\section{Discussion}\label{Discussion}

We here discuss whether and how our results are affected by the choice of the data analysis method. We then discuss the possible origin of the X-ray emission in GPs.

\subsection{Impact of Data Analysis Method}

Previous studies of X-ray luminosity in star-forming galaxies 
and its relation to SFR and metallicity probed various galaxy types. 
Among them, GPs most closely resemble in their extreme form. LBAs were selected as compact UV sources, which can represent a compact galaxy or a compact star-forming knot in a larger galaxy \citep{Overzier09}. In contrast, GPs are compact in the optical light as well. The similarity between GPs and some of the LBAs has been confirmed in our X-ray study: both samples occupy similar regions of our plots. 
It is therefore of extreme importance to verify the compatibility of the data analysis methods, especially between these two samples. We discuss here our choice of the SFR and metallicity estimators.

\subsubsection{SFR estimates} 
\label{discussion-sfr}

\citet{Brorby2016} determined the SFR for their LBAs from a combination of near-UV and infrared data. We instead adopted the H$\alpha$ SFR from \citet{Cardamone2009}.
Here, we test the impact of the 
different SFR determinations on our plots. We explore existing conversions
between H$\alpha,$ ultraviolet and infrared SFR estimators. 
  
\citet{Hunter2010} derived SFRs of low-metallicity, dust-poor blue compact dwarf galaxies (BCDs) both from H$\alpha$ and FUV. They obtained the ratio 
$\log {\rm SFR}_{\rm FUV}/\log {\rm SFR}_{\rm H\alpha} = 0.99$,
i.e. both SFR measurements provided comparable results.
\citet{Brorby2014} then studied  
the relation between the FUV SFR and that from the combined NUV+IR measurements for the same sample. They obtained 
${{\rm SFR}_{\rm NUV+IR}}/{{\rm SFR}_{\rm FUV}} = 1.23.$ 
Green Peas are similarly dust-poor as the BCDs and are UV bright; it is therefore  reasonable to assume that the UV fraction reprocessed and re-emitted by dust in the IR is low. For this reason, we apply the same conversions to our GP data to test their impact on the results. We first apply the H$\alpha$-to-FUV correction following \citet{Hunter2010}, and subsequently the FUV-to-NUV correction following \citet{Brorby2014}. 
The resulting SFRs are 
larger than the original \citet{Cardamone2009} values:
$70.3 \ M_\odot$\,yr$^{-1}$ for GP1, $44.9 \ M_\odot$\,yr$^{-1}$ for GP2,
and $22.7 \ M_\odot$\,yr$^{-1}$ for GP3 (see the values in Table~\ref{SFR_table}). 

Independently, \citet{Izotov2011} provided 
scaling factors between the FUV and H$\alpha$ SFRs for various sub-samples of their 
luminous compact star-forming galaxies overlapping with GPs. Their sub-samples were defined by the H$\beta$ equivalent width and morphology; we therefore measured the H$\beta$ equivalent widths for our three GPs from the SDSS spectra and assumed non-extended morphologies: EW(H$\beta$) = 90\,\AA\ for GP1, 200\,\AA\ for GP2, and 250\,\AA\ for GP3. 
The corresponding conversion factors from \citet{Izotov2011} (their Table~1) are: ${\rm SFR}_{\rm FUV}/{\rm SFR}_{\rm H\alpha} = 0.93$ for GP1, and ${\rm SFR}_{\rm FUV}/{\rm SFR}_{\rm H\alpha} = 0.68$ for GP2 and GP3. 
Applying this H$\alpha$-to-FUV conversion and then the FUV-to-NUV+IR 
conversion \citep{Brorby2014} as in the previous paragraph,
we obtain SFRs:
$67.3 \ M_\odot$\,yr$^{-1}$ for GP1, $31.3 \ M_\odot$\,yr$^{-1}$ for GP2,
and $15.7 \ M_\odot$\,yr$^{-1}$ for GP3. 

We present the results in Figure~\ref{SFR_comparison}, showing how the positions of the three GPs shift in the \citet{Brorby2016} diagram using the different SFR estimates. 
The conversion using the BCD fit predicts the highest SFR values 
and shifts GP1 and GP2 closer to the empirical law derived by \citet{Brorby2016}. 
Nevertheless, the X-ray excess is present in GP1 and GP2 for all of the conversions. 
It is unlikely that the SFR would be seriously underestimated by neglecting the dust re-emission (which would push the GPs to the right in Figure~\ref{brorby}, toward the SFR-metallicity plane), due to their dust-poor nature.  
The applied conversions thus show that the H$\alpha$ SFR applied in Section~\ref{results} does not affect our main conclusions.

\subsubsection{Metallicity Estimates} 
\label{discussion-metallicity}

Similar to SFR, different metallicity calibrations can affect the position of GPs in Figure~\ref{brorby} and in the right panel of Figure~\ref{basuzych}. 
We determined the GP metallicity using the 
[O\,{\sc iii}] and [N\,{\sc ii}] optical line ratios \citep[O3N2 method;][]{Pettini2004}, consistently with the control sample.
We checked that these values are in perfect agreement with the values obtained by \citet{Izotov2011} who used the so-called direct method, which is based on the electron temperature determination, $T_{\rm e},$ from the O$^{2+}$ line ratio, [O\,{\sc iii}]$\lambda4363/\left(\lambda4959+\lambda5007\right)$.

These 12+log(O/H) metallicities are $\sim\!0.5-0.7$\,dex lower than those reported by \citet{Cardamone2009}, the optical [O\,{\sc ii}] and [N\,{\sc ii}] lines \citep{Kewley2002}, employing the method of \citet{Tremonti2004}. 
The offset in metallicity from \citet{Cardamone2009} was also previously reported by \citet{Amorin2010} who used the direct method as well as the N2 method (where N2\,=\, log\,([\ion{N}{2}]$ \lambda 6584$/H{$\alpha$}).
\citet{Amorin2010} attributed the difference to the larger N/O ratios in the GPs than in other  star-forming galaxies.
However, \citet{Hawley2012} simply explained the offset of the \citet{Cardamone2009} results as due to the low metallicity of GPs, at which the O2N2 method is not applicable (12\,+\,log\,(O/H)\,$ \lesssim\!8.3$). Conversely, the O3N2 index of
\citet{Pettini2004} should be reliable in GPs, as their metallicity is above the lower limit set by \citet{Lopez-Sanchez2010}: $12+\log\rm{(O/H)}\!>\!8.0.$ 
\citet{Hawley2012} showed that the N2 method is valid for even  lower-metallicity galaxies and that the O3N2 method gives coincidentally correct estimates  due to the tight correlation with the N2 method.
The difference between our metallicity estimations and the O3N2 metallicity estimations derived from the line intensities by \citet{Hawley2012} for our GPs is less than $\sim\!1\%$, using both the original and \citet{Hawley2012} calibrations.

All the methods discussed in this paragraph,
except for that of \citet{Cardamone2009},
provide consistent metallicity results. 
Nevertheless, if we adopted the higher \citet{Cardamone2009} values, the data points would shift to the left in the
diagram of Figure~\ref{brorby}, making the X-ray flux of GPs even more enhanced with respect to other star-forming galaxies. 
Our main conclusions about the elevated X-ray flux of GP1 and GP2 are thus clearly unaffected by the metallicity measurements.



\subsection{Significance of the X-Ray Detection}
\label{xray_significance}
The significance of the GP1 and GP2 detection can be expressed by the likelihood parameter described in \citet{Watson2009}. It is based on a calculation of the probability that the detected number of counts is due to the Poissonian fluctuation of the background counts. The higher likelihood parameter corresponds to the lower probability of a spurious detection. Sources presented in the {\xmm} catalog 
reach a likelihood parameter of 6 or more. For a likelihood parameter equal to 10, the probability of a false detection is lower than 0.5\% \citep{Watson2009} and a likelihood parameter equal to 15 roughly corresponds to a 5$\sigma$ detection \citep{Cash1979}. Our GP1 and GP2 detections with the likelihood parameters 45 and 63, respectively, largely exceed the recommended thresholds.

Still, there is the possibility of an accidental detection of a background AGN 
that could contaminate the X-ray measurements.
Based on a deep extragalactic survey, we can estimate the probability of the presence of a source with a flux equal to the sensitivity limit of our observations. To make this estimate, we generated the sensitivity map of GP1 observation using the SAS tool {\sc{esensmap}}. The measured count rate in the sensitivity map was $\approx 0.0014$\,cts\,s$^{-1}$, corresponding to a flux $S = 10^{-15}$\,erg\,cm$^{-2}$\,s$^{-1}$. According to the deep {\xmm} survey of the 13$^H$ deep field by \citet{Loaring2005}, there should be $N(>S) = 900$ sources exceeding the flux $S$ in 1\,deg$^2$ for the energy interval 0.5-2\,keV (see their Fig.~8). A similar estimate can be also found in \citet{Mateos2008}, who reported a ${\log}N-{\log}S$ distribution for a large set of {\xmm} observations of sources with $\left|b\right|>20$\,deg. Scaling this number to the source extraction area, we get $N(>S)=0.2$ and $N(>S)=0.13$ for an extraction region with a radius of 30 $\arcsec$ (i.e., $0.00022$ deg$^2$), and 24 $\arcsec$ (i.e., $0.00014$ deg$^2$), respectively. Therefore, the probability of any significant contamination is quite low.

A similar argument can be used to estimate the expected number of background sources with the flux equal to the measured X-ray fluxes of GPs, i.e. to constrain the detection probability of a random background AGN instead of the studied source. For the GP1 flux (${\log}F_{x} \approx -14.7$), $N(>S) \approx 474$ in 0.5-2\,keV in 1 deg$^2$ according to \citet{Mateos2008}. This corresponds to $N(>\!S)=0.07$ for an extraction region with a radius of 24 $\arcsec$ (i.e., $0.00014$ deg$^2$). For  the GP2 flux (${\log}F_{x} \approx -14.4$), $N(>\!S) \approx 287$ in 0.5-2\,keV in $1$ deg$^2$, which corresponds to $N(>S)=0.06$ for an extraction region with a radius of 30 $\arcsec$ (i.e., $0.00022$ deg$^2$).
In addition, there is no indication from the X-ray images that there is 
any significant X-ray source present in the extraction regions other than the sources at the exact nominal positions of the GP galaxies. 

There is only a nearby X-ray source at the distance of 34 $\arcsec$ from GP1. For this reason, the radius of the extraction region of GP1 was reduced from 30 to 24 $\arcsec$ to eliminate any possible contamination.
According to the NED database, this source corresponds to the galaxy SDSS J074936.42+333641.5 with the optical magnitude 23 in the g-filter. Its X-ray spectrum is hard; it appears in the image only above 1\,keV and it is clearly detected in the hard 2-10\,keV band. This suggests that this source might be an obscured type-2 active galaxy, but the analysis of its X-ray spectrum is beyond the scope of this paper. Nevertheless, the possible contamination of our GP1 X-ray measurements by this source is negligible, as the GP1 source is not significantly detected in the hard band where the source has the highest flux. We do not expect any issues related to the source confusion (as this is occurring only for long exposures; \citet{Valtchanov2001}), or blending of multiple sources, which is an issue for overcrowded fields only.

\subsection{Stochasticity of the Results}
\label{stochasticity}

The excess in the X-ray emission observed in GP1 and GP2 with respect to other local star-forming galaxies suggests that a standard XRB population is not enough to account for their X-ray emission. 
In order to calculate the probability that the measured X-ray luminosity in our GPs is a statistical fluctuation around the mean value predicted by \citet{Brorby2016}, a detailed model for the total luminosity of XRBs in highly star-forming galaxies that includes the dependence on metallicity and that accounts for the variation of the XRB feedback due to statistical effects would be needed \citep[see, e.g., the model for the number of HMXBs in][]
{Douna2015}. Although the construction of such a model is not the main goal of this paper, before interpreting the X-ray excess by possible physical scenarios, we discuss here how likely the effects of stochasticity would be if the X-ray excess in our GPs was due to their standard XRB population.

It has been previously shown that for a randomly chosen galaxy with a given SFR, the total X-ray luminosity of a number of discrete sources can deviate from the expected value \citep[e.g.][]{Gilfanov2004b,Justham2012}. This effect is pronounced mainly in the low-SFR regime due to the strong asymmetry of the total luminosity probability distribution (in the non-linear low-SFR regime). On the other hand, for SFR $\gtrsim\!10$\,$M_\odot$\,yr$^{-1}$, the probability distribution is symmetric and becomes narrower with the higher SFR. 
\citet{Gilfanov2004b} showed the probability distribution of the total luminosity for the case of SFR\,=\,40\,$M_\odot$\,yr$^{-1}$, which is already so narrow that the probability of obtaining two times larger luminosity is lower than $p=0.001$ (see their Figure~6).
The SFR of our GP1 and GP2 is comparable or even higher, which would imply an even lower probability that the X-ray excess is only a statistical fluctuation from the distribution.

A similar result was obtained by \citet{Justham2012}. In addition, they showed the relative fraction and the cumulative distribution of the total X-ray luminosities for different SFRs. For   SFR\,$>\!10$\,$M_\odot$\,yr$^{-1}$, the cumulative distribution saturates at 1 for luminosities below 
$3~\langle L_{\mathrm x}\rangle$. The probability of detecting two out of three sources with the luminosity $4-6~\langle L_{\rm x} \rangle$, which is the case of our three GPs, is thus completely negligible.
We note that these estimates were done with 
models that do not include the metallicity dependence. However, as \citet{Justham2012} argue, 
the predicted increase in the number of sources at low-metallicity environments would imply that the spread in luminosity would be even lower in metal-poor environments. Considering the high SFR and low metallicity of the GPs, the measured X-ray fluxes are not due to stochastic deviations from the expected mean. Nevertheless, to precisely assess the likelihood of the observed luminosities, given the $L_{\rm x}$--SFR--metallicity plane of \citet{Brorby2016}, a robust statistical model for the aforementioned linear relation based on non-standard fitting techniques would be needed \citep[see, e.g., ][]{Gilfanov2004b,Douna2015}.

\subsection{Origin of the Enhanced X-Ray Emission}

We showed in previous Sections that the X-ray excess of the two detected GPs (GP1, GP2) cannot be affected by an incorrect placement of the sources in the diagram due to the SFR or metallicity determination methods, and that the excess is significant.
We note, however, that the $L_\mathrm{x}$--SFR--metallicity plane is based on 
a heterogeneous sample of galaxies of distinct types 
and the slope of the correlation might be strongly dependent on the observational sample \citep[see also][]{Douna2015,Brorby2016}.
Larger galaxy samples would be useful to populate the plane and check the robustness of its slope. Nevertheless, 
the location of GP1 and GP2 above the plane 
and the actual order of magnitude of their $L_\mathrm{x}$ 
suggests the need to further investigate the properties of their X-ray emission in comparison to other galaxies. 
We discuss here the possible explanations, while bearing in mind the low statistical weight of this GP sample.

\subsubsection{X-Ray Binary Populations and Stellar Evolution} 
\label{discussion_xrb_hotgas}

The X-ray emission of galaxies with a recent star-formation activity is proportional to the number of relatively short-lived HMXBs, and thus to the SFR. \citet{Mineo2014} provided the relation $L_{0.5-8\,\rm{keV}} \, [{\rm erg \ s^{-1}}] \approx 4 \times 10^{39} \, {\rm SFR \ [M_{\odot} \ yr^{-1}]}$. The contribution of the older, long-lived, LMXBs is instead proportional to the stellar mass,
$L_{\rm LMXB, 0.5-8\,keV} \, ({\rm erg \ s^{-1}}) \approx 10^{39}\times \frac{\rm{M}_*}{10^{10}\rm{M}_\odot}$ \citep{Gilfanov2014}, and the X-ray luminosity function has a cutoff at $\log L_\mathrm{x} \mathrm{[erg\,s^{-1}]}\approx 39-39.5$ \citep{Gilfanov2004a}. Therefore, the LMXBs do not significantly contribute to the total X-ray emission when SFR\,$>\,$1\,$M_{\odot}$\,yr$^{-1}$ \citep{Gilfanov2004a}. For our GP sources with SFR~$>\!10$\,$M_{\odot}$\,yr$^{-1}$, $\log L_\mathrm{x}\mathrm{[erg\,s^{-1}]} \approx 42$, and $\log {\rm{M}}_*\mathrm{[\rm{M}_\odot]}\!<\!10$, the contribution of LMXBs is insignificant ($<\!1\%$). This is also indicated by the vertical line in Figure~\ref{basuzych} at the limit $\log$(sSFR\,[yr$^{-1}$])\,$ \equiv$ -10, distinguishing between the regions of HMXB and LMXB dominance. GPs with $\log$(sSFR\,[yr$^{-1}$])\,=\,-8 are clearly HMXB-dominated and the
large X-ray enhancement cannot be attributed to the LMXBs.

The traditional L$_\mathrm{x}$--SFR relations underestimated the X-ray emission in this regime, hence the necessity to include metallicity. The HMXB formation is more efficient in metal-poor galaxies, as shown by, e.g., \citet{Douna2015} and \citet{Basu-Zych2016}. We used the metallicity-dependent L$_\mathrm{x}$-SFR relation from \citet{Brorby2016} for the comparison of GPs with other star-forming galaxies (Figure~\ref{brorby}). However, the low GP metallicity was not sufficient to explain the observed L$_\mathrm{x}\!\sim\!10^{42}$\,erg\,s$^{-1}$ emission in GP1 and GP2, which is still  a factor of $4-6$ larger than the prediction. Nonetheless, the relation of \citet{Brorby2016} was established in a purely empirical method, based on the available data, and its slope may need to be revised with new observations. 

Alternatively, some authors have speculated about the deviation of the initial mass function (IMF), such as 
the top-heavy IMF that might be more relevant for compact dwarf galaxies 
\citep{Dabringhausen2009, Dabringhausen2012, Marks2012, Stanway2016}.
The IMF shape has a direct impact on the number and luminosity of the black hole systems, in the form of both HMXBs and LMXBs, and on the supernova rate \citep[see also the discussion in][]{Justham2012}. 
Indirectly, it also enters the SFR derivation (the Kennicutt relation implicitly assumes the Salpeter IMF).
The theoretical considerations of the IMF are potentially interesting for the GPs, but we have no means of testing the models with the present data.

Empirically, in the cases of an unexplained bright X-ray emission 
\citep[as in LBAs;][]{Basu-Zych2013a, Prestwich2015},  
authors commonly report the presence of off-nuclear, point-like sources with 
luminosities, $L_\mathrm{x}\!>\!10^{39}$\,erg\,s$^{-1},$ known as ultra-luminous X-ray sources
(ULXs). Their interpretations vary from stellar-mass compact objects in HMXBs accreting at super-Eddington rates or with anisotropic or relativistically beamed emission \citep{Bachetti2014}, to intermediate-mass black holes (IMBHs), or even a combination of both. For a review of ULXs, we refer to \citet{Kaaret2017a} and references therein. 
\citet{Basu-Zych2013a} claimed to detect a higher number of off-nuclear ULXs per unit SFR than predicted theoretically in several spatially resolved galaxies, and cite other works with similar examples. The overabundance of ULXs is 
mainly associated with low-metallicity environments \citep[e.g.][and references therein]{Mapelli2010,Basu-Zych2016}. 
Our \xmm\ GP data lack spatial resolution, therefore we can only speculate about the X-ray source distribution.

The X-ray emission from stellar-mass compact objects seems to be a promising avenue for explaining the high $L_\mathrm{x}$ in GP1 and GP2. However,
it is not clear why there is such a large difference between X-ray properties in otherwise similar galaxies -- GP1 and GP2 on one side, and GP3 with the GP analogs on the other.
The presence (or absence) of several ULXs or accreting IMBHs would provide a solution. Their powerful X-ray output would help explain the excess observed in GP1 and GP2. Their small numbers and the short lifetimes of the X-ray phase would account for the variations between the targets.     
Resolved observations are necessary for testing the different scenarios.

\subsubsection{Hot Gas Emission}

Star forming galaxies possess large amounts of hot gas, which is yet another source of X-ray photons, mainly on sub-keV energies. The gas is mostly located in the star-forming regions and is assumed to be in an outflowing state, driven by stellar winds and supernovae; its emission is therefore proportional to the SFR. In galaxies with high signal-to-noise ratio X-ray spectra, the hot gas, thermal component is directly identified, superposed on a power law \citep[e.g.][]{Oti-Floranes2012,Oti-Floranes2014}.
\citet{Mineo2012b} estimated that the hot gas contributes on average 
$\sim\!30\%$ to the total X-ray emission in star-forming galaxies, with potentially large variations. They reported the relation between the diffuse X-ray luminosity from hot gas and SFR as $L^{\rm diff}_{\rm 0.5-2\,keV} \approx 8.3\times10^{38}$ SFR.

The measured steep spectral slope of GP1 and its non-detection in the hard X-ray band may suggest a significant contribution in the soft X-rays that could be modeled as plasma emission due to hot gas. However, the data quality does not allow us to discriminate between a simple power-law model and a  hot plasma emission model, or a combination of both. It is impossible to determine the hot gas contribution in any of our GPs with the current data. Nevertheless, if we considered the above-mentioned relation from \citet{Mineo2012b}, the expected X-ray luminosity from hot gas would be $L^{\rm diff}_{\rm 0.5-2\,keV} \approx 5\times10^{40}$\,erg\,s$^{-1}$ for GP1 (with SFR\,=\,58.8\,$M_\odot$\,yr$^{-1}$). This corresponds to less than 5\% of the measured X-ray luminosity, and it is thus unlikely to explain the observed X-ray excess by the hot gas emission.

\subsubsection{Possible AGN Contribution} \label{sec_AGN}

The presence of an AGN would be another possible explanation for the elevated X-ray emission. Based on the measured X-ray luminosity, we can estimate the accretion rate with respect to the Eddington luminosity. For a supposed black hole with mass $\sim\!\!10^5$\,$M_\odot$ \citep[see the typical masses of central black holes in dwarf galaxies in, e.g.][]{Mezcua2018}, the Eddington luminosity is of the order of $10^{43}$\,erg\,s$^{-1}$. The excess of $\sim\!10^{42}$\,erg\,s$^{-1}$ would correspond to an accretion rate of $\dot{m} \approx\!0.1 \dot{m}_{\rm Edd}$. 
This accretion rate is typical of Seyfert galaxies,
which often exhibit an optical/UV excess.  
However, if the excess is due to the thermal emission from a black hole accretion disk and if the mass of the central black hole is small (such as expected in GPs), the excess shifts toward high energies -- the peak temperature would be at $\sim\!$ 0.3\,keV (0.1\,keV) for a highly spinning (non-rotating) black hole with mass 
$\sim\!10^5$\,$M_\odot$ \citep[see, e.g.,][]{Svoboda2017}. Therefore, the accretion-disk thermal emission would fall to soft X-rays and to the band between the FUV and X-rays. For lower accretion efficiencies,
the total AGN luminosity is lower and the radiation is typically harder,  hence it would  be even more difficult to detect direct AGN signatures in other than the X-ray band.
Only when inefficient accretion leads to a jet production, which is typical, e.g., of low-ionization nuclear emission-line region (LINER) galaxies \citep{Ho2008},
significant radio emission can be produced in the jet. However, the radio emission of our GPs was found to be even dimmer than would be expected from the star formation \citep{Chakraborti2012}.

\begin{figure}[tb!]
\includegraphics[width=0.49\textwidth]{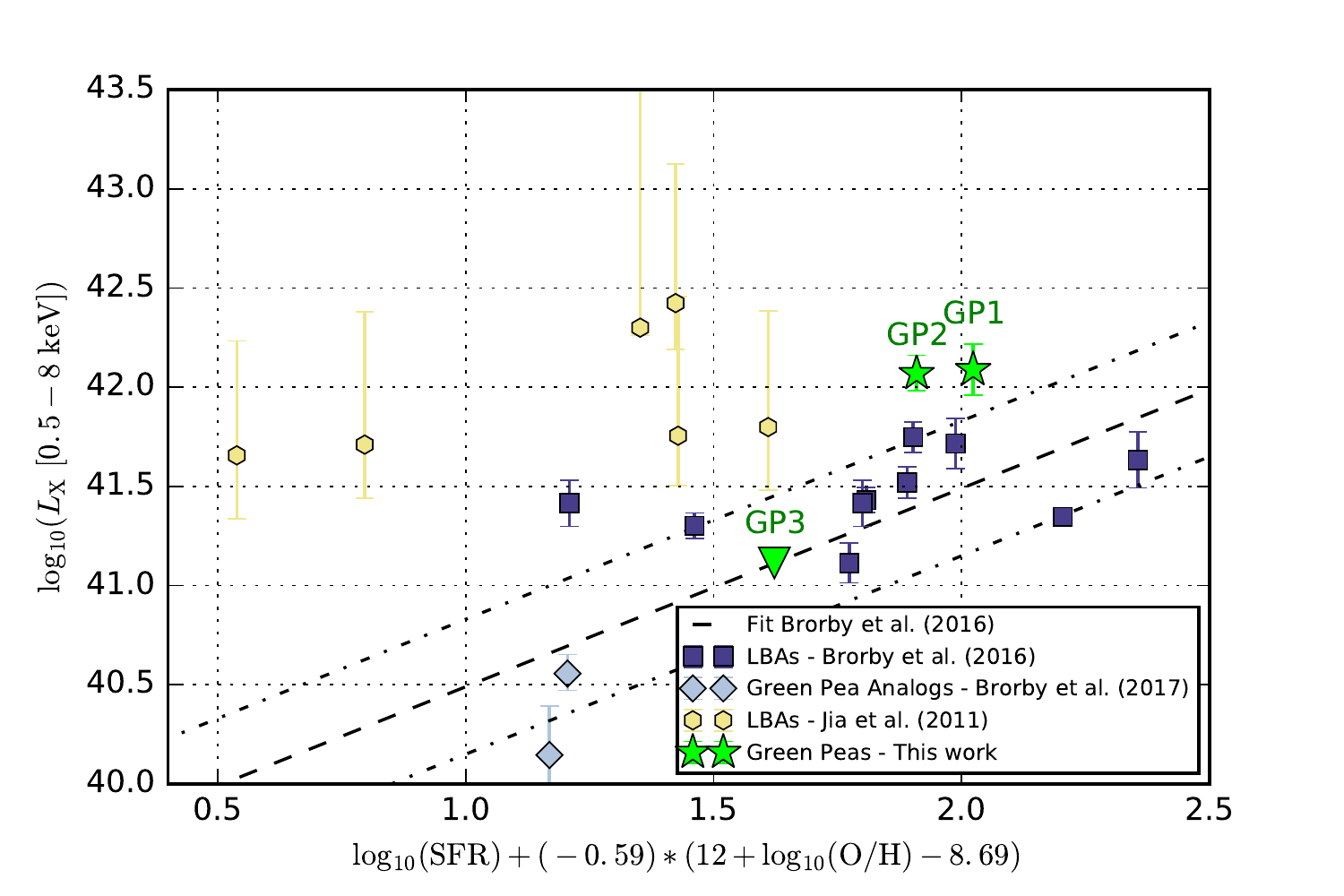}
\caption{Comparison of GP galaxies with composite LBAs that are expected to contain an obscured active nucleus \citep[sample of][yellow circles]{Jia2011}.
}
\label{Jia_comparison}
\end{figure}

The X-ray luminosity of GP1 and GP2 places them in the region of the L$_{x}$--SFR--metallicity plane where AGNs could be expected.
Figure~\ref{Jia_comparison} shows an X-ray comparison between the GP galaxies and composite LBAs from the sample of \citet{Jia2011} that were argued
to contain an obscured active nucleus, unlike the LBAs from \citet{Brorby2016}. The two LBA samples occupy very different regions of the BPT classification diagram (see Figure~\ref{fig_bpt}). To plot the composite LBAs in the Brorby diagram,
we used the H$\alpha$ SFRs and the O3N2 oxygen abundances derived by \citet{Overzier09}.
Their X-ray luminosities in the 0.5-8\,keV energy band were obtained by extrapolating the $L_{\rm 0.5-2\,keV}$ values reported in Table~4 of \citet{Jia2011}, while assuming an X-ray power law with the slope $\Gamma = 1.9$.  Although the error bars of the measurements are relatively large due to low exposure times (10\,-\,20\,ks), there is an evident X-ray excess with respect to the trend of LBA sources that were selected to be purely star-forming with only a limited probability of the AGN occurrence by \citet{Brorby2016}. The level of the X-ray luminosity of LBAs with supposed obscured AGNs is of the same order as the detected X-ray luminosity of GP1 and GP2, suggesting that a hidden AGN can indeed be a viable scenario for the measured X-ray excess in GP1 and GP2.

The optical SDSS emission line ratios for all of our three GPs are compatible with pure star-formation, albeit in an extreme form.
The GPs are situated on the star-forming sequence of the BPT diagram (Figure~\ref{fig_bpt}), unlike the \citet{Jia2011} galaxies. Nevertheless, they lie near the borderline that separates star-forming galaxies from AGNs, which could raise questions about the AGN contribution.  
The BPT diagram separation curve was derived by \citet{Kewley2001} who simulated gas emission in the limit of extreme starburst conditions, while considering realistic stellar and ISM parameters.
However, \citet{Kewley13,Kewley13b} later showed (and verified by observation) that the classification line shifts upward in the conditions encountered mainly in $z>1$ galaxies, but possibly in GPs as well:
large ionization parameter, hard radiation field and/or high densities of  ionized gas. 
If the shifted limit is applicable to GPs,
they will be safely classified as star-forming galaxies. 
High ionization parameters and high densities of ionized gas have also been confirmed locally, in the star-forming clumps of M82 and the Antennae galaxies \citep{Smith2006a,Snijders2007}. 
Extreme conditions cannot be ruled out in the GPs, as witnessed by their 
optical emission-line ratios, more similar to high-$z$ galaxies than to the local samples \citep{Schaerer16}. In conclusion, the BPT diagram alone does not provide any direct indication of an AGN in our GP sample. 

As for other commonly used optical signatures, the emission line widths do not show signatures of an AGN broad-line region in any of our targets. Nevertheless, as the X-ray spectral index of GP1 is large and is similar to that encountered in narrow-line Seyfert\,1 nuclei \citep{Boller1996, Done2012}, 
we have checked for the presence of \ion{Fe}{10} lines, which are commonly found in such objects \citep{Mullaney08}. We detected a weak feature in the SDSS spectrum which may, however, be an artifact. A deeper exposure would be needed to confirm its relevance.


\subsection{A Possible Link between X-Rays and the LyC Leakage?}


The SFR measurement in this paper
was based on the H$\alpha$ line flux, assuming that all of the flux was captured by the SDSS aperture and that 
all of the ionizing radiation was reprocessed by the interstellar gas. In contrast, if there were an ionizing flux leakage from the galaxy, the derived SFR would be underestimated \citep{Izotov2016b}. A LyC escape from several GP-like galaxies has indeed been observed \citep{Izotov2016,Izotov2016b,Izotov2018a,Izotov2018b}, with escape fractions ranging from $\sim\!\!5\%$ to $\sim\!\!70\%.$
If the X-ray excess in GP1 and GP2 were due to an underestimated H$\alpha$ SFR measurement, the real SFR of GP1 could be as high as $300$\,$M_{\odot}$\,yr$^{-1},$ as we derived from X-rays in previous sections (and analogously for the remaining two galaxies). 
SFR estimations from the UV, IR and radio fluxes and their possible discrepancies would therefore be useful to clarify the issue, 
even though the general consistency between the H$\alpha$-derived and the FUV-derived SFRs in GPs \citep{Izotov2011} indicates that this effect cannot account for the too high X-ray emission in two of our targets.

No direct LyC observations are currently available for the targets studied in this paper but, as we know from the optical and UV data, the GP galaxies have a vigorous, on-going star-formation activity, leading to the formation of black hole systems as well as to the ionization of the ISM. X-ray binaries are the final stages of stellar evolution and are thus the evidence of recent presence of massive stars providing the necessary ionizing photons. The X-ray-emitting systems provide energy capable of modifying the 
surrounding medium. 
First, the X-ray photons heat and ionize the 
gas and destroy the dust particles along their path. Second, the production of X-rays is associated with energetic processes, and probes the presence of copious UV radiation and high-energy particles that modify the thermal and ionization state of the ISM. Third, the X-ray-emitting systems are sources of powerful jets and/or outflows. Their mechanical energy 
heats the ISM and can remove gas and dust along the direction of propagation. 
Among others, these processes may facilitate the escape of LyC 
photons, perhaps along preferential channels. 
The role of feedback processes for the LyC escape has been debated in the 
literature for a long time: for instance, \citet{Heckman2011} speculated about the role of fast outflows;
\citet{Alexandroff2015} proposed the SFR surface density (i.e. SFR/kpc$^2$)
as the main parameter to characterize the efficiency of creating ``holes'' in the 
ISM. \citet{Verhamme2017} provided the first direct confirmation that the LyC escape  fraction from GP-like galaxies indeed correlates with SFR/kpc$^2$.  
 
X-ray observations of confirmed LyC leakers are available for Haro\,11 \citep{Basu-Zych2016} and Tol\,1247-232 \citep{Kaaret2017}. The LyC escape fractions are low in these two targets, on the level of $<\!5\%$ \citep{Chisholm2017}. 
\citet{Basu-Zych2016} worked with the hypothesis that one of the Haro\,11 star-forming knots harbours a hidden AGN, suggested by independent multi-wavelength observations. Removing the possible AGN contribution, they still found an excess of bright X-ray sources. \citet{Prestwich2015} proposed an IMBH as a potential candidate for this X-ray excess. For Tol\,1247-232, \citet{Kaaret2017} did not rule out a low-luminosity AGN contribution despite the absence of its signature in the optical spectra, and also speculated about the presence of a few ULXs or one hyper-luminous X-ray source (HLX). 
The HLX could either be a source accreting at a highly super-Eddington rate or an IMBH.
They made a connection between an X-ray source and a feedback feature, observed in
H$\alpha$ \citep{Puschnig2017} and possibly related to the LyC escape. 

Larger samples of X-ray and multi-wavelength observations for confirmed LyC leakers are necessary to understand the nature and role of their X-ray sources. 
The studies that resolved individual X-ray sources in nearby LyC leakers systematically speculate about the presence of AGNs. If AGNs were confirmed in GPs, including those with a LyC escape, 
this would have profound implications for the interpretation of the LyC escape mechanisms and the cosmic reionization.

\section{Conclusions}
\label{Conclusions}

In this paper, we have analyzed the first X-ray observations of low-mass, metal-poor, star-forming galaxies known as the Green Peas, obtained with the {\xmm} satellite. We have detected X-ray flux in two targets (GP1 and GP2) and have put an upper limit on the third target (GP3).
The X-ray luminosities of the two detected targets are of the order of $\sim\! 10^{42}$\,erg\,s$^{-1}$, and are a factor of $4-6$ larger than predicted by empirical laws connecting the X-ray luminosity of star-forming galaxies to their SFR and metallicity. This suggests that the two Green Pea galaxies may either (i) have an additional source of X-ray flux (hidden AGN, IMBH, or ULX), or (ii) have a larger and/or brighter population of HMXBs than usual star-forming galaxies, or (iii) their SFR from the optical spectral lines is largely underestimated. 

While it is impossible to uniquely determine the physical origin of the observed X-ray excess with the current data,
the feasible scenarios should explain both the level of the X-ray excess detected in GP1 and GP2, as well as no X-ray excess in GP3.
The hidden AGN scenario would be consistent with the X-ray measurements and can easily explain a scatter in the X-ray luminosity of different GPs (the activity is either on or off), but this scenario is not consistent with the optical diagnostics, indicating that the dominant GP flux comes from star formation. Pure star formation with a standard X-ray binary population does not seem to be capable of producing such a large amount of X-rays, not even after including the contribution of hot gas. The missing X-ray luminosity could be supplied by several ULXs. However, the presence of ULXs is impossible to confirm without spatially resolved X-ray images. The large X-ray luminosity could alternatively be explained by a modified IMF that would predict larger numbers of X-ray binaries in the star-forming galaxies. However, if the modified IMF was universal for the star-forming dwarf galaxies, it fails to explain why some of the galaxies, such as GP3, are faint in the X-ray. 

The detection of powerful X-ray emission in two out of the three GPs adds a piece to the puzzle of trying to understand the nature of the GPs and their relation to the local and distant galaxies: their mode of star formation, their ISM structure, and their ionizing LyC escape. The X-ray sources are probes of the high-mass objects and of the energetic processes supplying radiative and mechanical energy to the ISM.  
To reach a better understanding of the link between the different processes, sensitive, spatially resolved X-ray observations of low-metallicity, dwarf, star-forming galaxies will be necessary, and will hopefully be possible in the future.


\acknowledgments
We thank the referee for their insightful comments that improved the paper quality. We thank Ma\"{i}ca Clavel for useful discussions about the X-ray data analysis. 
J.S. and I.O. acknowledge financial support from the Czech Science Foundation project No.17-06217Y. 
I.O. is grateful to Universit\'e Grenoble Alpes IPAG for their hospitality during the paper preparation. 
We have made use of the SDSS, ADS, and NED databases.




\bibliographystyle{aa}
\bibliography{/home/jirka/Documents/references}

\begin{thebibliography}{131}
\expandafter\ifx\csname natexlab\endcsname\relax\def\natexlab#1{#1}\fi

\bibitem[{Ade {et~al.}(2016)Ade, Aghanim, Arnaud, Ashdown, Aumont, Baccigalupi,
  Banday, Barreiro, Bartlett, Bartolo, Battaner, Battye, Benabed, Beno{\^{i}}t,
  Benoit-L{\'{e}}vy, Bernard, Bersanelli, Bielewicz, Bock, Bonaldi, Bonavera,
  Bond, Borrill, Bouchet, Boulanger, Bucher, Burigana, Butler, Calabrese,
  Cardoso, Catalano, Challinor, Chamballu, Chary, Chiang, Chluba, Christensen,
  Church, Clements, Colombi, Colombo, Combet, Coulais, Crill, Curto, Cuttaia,
  Danese, Davies, Davis, de~Bernardis, de~Rosa, de~Zotti, Delabrouille,
  D{\'{e}}sert, {Di Valentino}, Dickinson, Diego, Dolag, Dole, Donzelli,
  Dor{\'{e}}, Douspis, Ducout, Dunkley, Dupac, Efstathiou, Elsner, En{\ss}lin,
  Eriksen, Farhang, Fergusson, Finelli, Forni, Frailis, Fraisse, Franceschi,
  Frejsel, Galeotta, Galli, Ganga, Gauthier, Gerbino, Ghosh, Giard,
  Giraud-H{\'{e}}raud, Giusarma, Gjerl{\o}w, Gonz{\'{a}}lez-Nuevo,
  G{\'{o}}rski, Gratton, Gregorio, Gruppuso, Gudmundsson, Hamann, Hansen,
  Hanson, Harrison, Helou, Henrot-Versill{\'{e}}, Hern{\'{a}}ndez-Monteagudo,
  Herranz, Hildebrandt, Hivon, Hobson, Holmes, Hornstrup, Hovest, Huang,
  Huffenberger, Hurier, Jaffe, Jaffe, Jones, Juvela, Keih{\"{a}}nen, Keskitalo,
  Kisner, Kneissl, Knoche, Knox, Kunz, Kurki-Suonio, Lagache,
  L{\"{a}}hteenm{\"{a}}ki, Lamarre, Lasenby, Lattanzi, Lawrence, Leahy,
  Leonardi, Lesgourgues, Levrier, Lewis, Liguori, Lilje, Linden-V{\o}rnle,
  L{\'{o}}pez-Caniego, Lubin, Mac{\'{i}}as-P{\'{e}}rez, Maggio, Maino,
  Mandolesi, Mangilli, Marchini, Maris, Martin, Martinelli,
  Mart{\'{i}}nez-Gonz{\'{a}}lez, Masi, Matarrese, McGehee, Meinhold,
  Melchiorri, Melin, Mendes, Mennella, Migliaccio, Millea, Mitra,
  Miville-Desch{\^{e}}nes, Moneti, Montier, Morgante, Mortlock, Moss, Munshi,
  Murphy, Naselsky, Nati, Natoli, Netterfield, N{\o}rgaard-Nielsen, Noviello,
  Novikov, Novikov, Oxborrow, Paci, Pagano, Pajot, Paladini, Paoletti,
  Partridge, Pasian, Patanchon, Pearson, Perdereau, Perotto, Perrotta,
  Pettorino, Piacentini, Piat, Pierpaoli, Pietrobon, Plaszczynski,
  Pointecouteau, Polenta, Popa, Pratt, Pr{\'{e}}zeau, Prunet, Puget, Rachen,
  Reach, Rebolo, Reinecke, Remazeilles, Renault, Renzi, Ristorcelli, Rocha,
  Rosset, Rossetti, Roudier, Rouill{\'{e}}~d'Orfeuil, Rowan-Robinson,
  Rubi{\~{n}}o-Mart{\'{i}}n, Rusholme, Said, Salvatelli, Salvati, Sandri,
  Santos, Savelainen, Savini, Scott, Seiffert, Serra, Shellard, Spencer,
  Spinelli, Stolyarov, Stompor, Sudiwala, Sunyaev, Sutton, Suur-Uski, Sygnet,
  Tauber, Terenzi, Toffolatti, Tomasi, Tristram, Trombetti, Tucci, Tuovinen,
  T{\"{u}}rler, Umana, Valenziano, Valiviita, {Van Tent}, Vielva, Villa, Wade,
  Wandelt, Wehus, White, White, Wilkinson, Yvon, Zacchei, \&
  Zonca}]{PlanckCollaboration2015}
Ade, P. A.~R., Aghanim, N., Arnaud, M., {et~al.} 2016, \aap, 594, A13

\bibitem[{Albareti {et~al.}(2017)Albareti, {Allende Prieto}, Almeida, Anders,
  Anderson, Andrews, Arag{\'{o}}n-Salamanca, Argudo-Fern{\'{a}}ndez, Armengaud,
  Aubourg, \& al.}]{Albareti17}
Albareti, F., {Allende Prieto}, C., Almeida, A., {et~al.} 2017, \apjs, 233, 25

\bibitem[{Alexandroff {et~al.}(2015)Alexandroff, Heckman, Borthakur, Overzier,
  \& Leitherer}]{Alexandroff2015}
Alexandroff, R.~M., Heckman, T.~M., Borthakur, S., Overzier, R., \& Leitherer,
  C. 2015, \apj, 810, 104

\bibitem[{Amor{\'{i}}n {et~al.}(2010)Amor{\'{i}}n, P{\'{e}}rez-Montero, \&
  V{\'{i}}lchez}]{Amorin2010}
Amor{\'{i}}n, R.~O., P{\'{e}}rez-Montero, E., \& V{\'{i}}lchez, J.~M. 2010,
  \apj Lett., 715, L128

\bibitem[{Arnaud(1996)}]{Arnaud1996}
Arnaud, K. 1996, Astron. Data Anal. Softw. Syst. V, 101, 17

\bibitem[{Artale {et~al.}(2015)Artale, Tissera, \& Pellizza}]{Artale2015}
Artale, M.~C., Tissera, P.~B., \& Pellizza, L.~J. 2015, \mnras, 448, 3071

\bibitem[{Bachetti {et~al.}(2014)Bachetti, Harrison, Walton, Grefenstette,
  Chakrabarty, F{\"{u}}rst, Barret, Beloborodov, Boggs, Christensen, Craig,
  Fabian, Hailey, Hornschemeier, Kaspi, Kulkarni, MacCarone, Miller, Rana,
  Stern, Tendulkar, Tomsick, Webb, \& Zhang}]{Bachetti2014}
Bachetti, M., Harrison, F.~A., Walton, D.~J., {et~al.} 2014, Nature, 514, 202

\bibitem[{Baldwin {et~al.}(1981)Baldwin, Phillips, \& Terlevich}]{Baldwin1981}
Baldwin, A., Phillips, M.~M., \& Terlevich, R. 1981, Publ. Astron. Soc.
  Pacific, 93, 5

\bibitem[{Barkana \& Loeb(2001)}]{Barkana2001}
Barkana, R. \& Loeb, A. 2001, Phys. Rep., 349, 125

\bibitem[{Basu-Zych {et~al.}(2016)Basu-Zych, Lehmer, Fragos, Hornschemeier,
  Yukita, Zezas, \& Ptak}]{Basu-Zych2016}
Basu-Zych, A.~R., Lehmer, B.~D., Fragos, T., {et~al.} 2016, \apj, 818, 140

\bibitem[{Basu-Zych {et~al.}(2013{\natexlab{a}})Basu-Zych, Lehmer,
  Hornschemeier, Bouwens, Fragos, Oesch, Belczynski, Brandt, Kalogera, Luo,
  Miller, Mullaney, Tzanavaris, Xue, \& Zezas}]{Basu-Zych2013a}
Basu-Zych, A.~R., Lehmer, B.~D., Hornschemeier, A.~E., {et~al.}
  2013{\natexlab{a}}, \apj, 762, 45

\bibitem[{Basu-Zych {et~al.}(2013{\natexlab{b}})Basu-Zych, Lehmer,
  Hornschemeier, Gon{\c{c}}alves, Fragos, Heckman, Overzier, Ptak, \&
  Schiminovich}]{Basu-Zych2013}
Basu-Zych, A.~R., Lehmer, B.~D., Hornschemeier, A.~E., {et~al.}
  2013{\natexlab{b}}, \apj, 774, 152

\bibitem[{Best \& Heckman(2012)}]{Best2012}
Best, P.~N. \& Heckman, T.~M. 2012, \mnras, 421, 1569

\bibitem[{Bian {et~al.}(2017)Bian, Fan, McGreer, Cai, \& Jiang}]{Bian2017}
Bian, F., Fan, X., McGreer, I., Cai, Z., \& Jiang, L. 2017, \apj, 837, L12

\bibitem[{Boller {et~al.}(1996)Boller, Brandt, \& Fink}]{Boller1996}
Boller, T., Brandt, W.~N., \& Fink, H. 1996, \aap, 305, 53

\bibitem[{Bouwens {et~al.}(2015{\natexlab{a}})Bouwens, Illingworth, Oesch,
  Trenti, Labb{\'{e}}, Bradley, Carollo, van Dokkum, Gonzalez, Holwerda, Franx,
  Spitler, Smit, \& Magee}]{Bouwens15}
Bouwens, R., Illingworth, G., Oesch, P., {et~al.} 2015{\natexlab{a}}, \apj,
  803, 34

\bibitem[{Bouwens {et~al.}(2015{\natexlab{b}})Bouwens, Illingworth, Oesch,
  Caruana, Holwerda, Smit, \& Wilkins}]{Bouwens2015}
Bouwens, R.~J., Illingworth, G.~D., Oesch, P.~A., {et~al.} 2015{\natexlab{b}},
  \apj, 811, 140

\bibitem[{Bowman {et~al.}(2018)Bowman, Rogers, Monsalve, Mozdzen, \&
  Mahesh}]{Bowman2018}
Bowman, J., Rogers, A., Monsalve, R., Mozdzen, T., \& Mahesh, N. 2018, \nat,
  555, 67

\bibitem[{Brinchmann {et~al.}(2004)Brinchmann, Charlot, White, Tremonti,
  Kauffmann, Heckman, \& Brinkmann}]{Brinchmann2004}
Brinchmann, J., Charlot, S., White, S. D.~M., {et~al.} 2004, \mnras, 351, 1151

\bibitem[{Brorby \& Kaaret(2017)}]{Brorby2017}
Brorby, M. \& Kaaret, P. 2017, \mnras, 470, 606

\bibitem[{Brorby {et~al.}(2014)Brorby, Kaaret, \& Prestwich}]{Brorby2014}
Brorby, M., Kaaret, P., \& Prestwich, A. 2014, \mnras, 441, 2346

\bibitem[{Brorby {et~al.}(2016)Brorby, Kaaret, Prestwich, \&
  Mirabel}]{Brorby2016}
Brorby, M., Kaaret, P., Prestwich, A., \& Mirabel, I.~F. 2016, \mnras, 457,
  4081

\bibitem[{Cardamone {et~al.}(2009)Cardamone, Schawinski, Sarzi, Bamford,
  Bennert, Urry, Lintott, Keel, Parejko, Nichol, Thomas, Andreescu, Murray,
  Raddick, Slosar, Szalay, \& Vandenberg}]{Cardamone2009}
Cardamone, C., Schawinski, K., Sarzi, M., {et~al.} 2009, \mnras, 399, 1191

\bibitem[{Cash(1976)}]{Cash1976}
Cash, W. 1976, \aap, 52, 307

\bibitem[{Cash(1979)}]{Cash1979}
Cash, W. 1979, \apj, 228, 939

\bibitem[{Chakraborti {et~al.}(2012)Chakraborti, Yadav, Cardamone, \&
  Ray}]{Chakraborti2012}
Chakraborti, S., Yadav, N., Cardamone, C., \& Ray, A. 2012, \apj, 746, L6

\bibitem[{Chisholm {et~al.}(2017)Chisholm, Orlitov{\'{a}}, Schaerer, Verhamme,
  Worseck, Izotov, Thuan, \& Guseva}]{Chisholm2017}
Chisholm, J., Orlitov{\'{a}}, I., Schaerer, D., {et~al.} 2017, \aap, 605, A67

\bibitem[{Choudhury \& Ferrara(2006)}]{Choudhury2006}
Choudhury, T.~R. \& Ferrara, A. 2006, \mnras Lett., 371, L55

\bibitem[{Colbert {et~al.}(2004)Colbert, Heckman, Ptak, Strickland, \&
  Weaver}]{Colbert2004}
Colbert, E. J.~M., Heckman, T.~M., Ptak, A.~F., Strickland, D.~K., \& Weaver,
  K.~A. 2004, \apj, 602, 231

\bibitem[{Dabringhausen {et~al.}(2009)Dabringhausen, Kroupa, \&
  Baumgardt}]{Dabringhausen2009}
Dabringhausen, J., Kroupa, P., \& Baumgardt, H. 2009, \mnras, 394, 1529

\bibitem[{Dabringhausen {et~al.}(2012)Dabringhausen, Kroupa, Pflamm-Altenburg,
  \& Mieske}]{Dabringhausen2012}
Dabringhausen, J., Kroupa, P., Pflamm-Altenburg, J., \& Mieske, S. 2012, \apj,
  747 [\eprint[arXiv]{1110.2779}]

\bibitem[{de~Barros {et~al.}(2016)de~Barros, Vanzella, Amor{\'{i}}n,
  Castellano, Siana, Grazian, Suh, Balestra, Vignali, Verhamme, Zamorani,
  Mignoli, Hasinger, Comastri, Pentericci, P{\'{e}}rez-Montero, Fontana,
  Giavalisco, \& Gilli}]{deBarros16}
de~Barros, S., Vanzella, E., Amor{\'{i}}n, R., {et~al.} 2016, \aap, 585, A51

\bibitem[{Done {et~al.}(2012)Done, Davis, Jin, Blaes, \& Ward}]{Done2012}
Done, C., Davis, S., Jin, C., Blaes, O., \& Ward, M. 2012, \mnras, 420, 1848

\bibitem[{Douna {et~al.}(2018)Douna, Pellizza, Laurent, \& Mirabel}]{Douna2017}
Douna, V.~M., Pellizza, L.~J., Laurent, P., \& Mirabel, I.~F. 2018, \mnras,
  474, 3488

\bibitem[{Douna {et~al.}(2015)Douna, Pellizza, Mirabel, \& Pedrosa}]{Douna2015}
Douna, V.~M., Pellizza, L.~J., Mirabel, I.~F., \& Pedrosa, S.~E. 2015, \aap,
  579, A44

\bibitem[{Fabbiano(2006)}]{Fabbiano06}
Fabbiano, G. 2006, \araa, 44, 323

\bibitem[{Faucher-Gigu{\`{e}}re {et~al.}(2008)Faucher-Gigu{\`{e}}re, Lidz,
  Hernquist, \& Zaldarriaga}]{FaucherGiguere2008}
Faucher-Gigu{\`{e}}re, C., Lidz, A., Hernquist, L., \& Zaldarriaga, M. 2008,
  \apj, 688, 85

\bibitem[{Fontanot {et~al.}(2012)Fontanot, Cristiani, \&
  Vanzella}]{Fontanot2012}
Fontanot, F., Cristiani, S., \& Vanzella, E. 2012, \mnras, 425, 1413

\bibitem[{Fragos {et~al.}(2013{\natexlab{a}})Fragos, Lehmer, Tremmel,
  Tzanavaris, Basu-Zych, Belczynski, Hornschemeier, Jenkins, Kalogera, Ptak, \&
  Zezas}]{Fragos2013}
Fragos, T., Lehmer, B., Tremmel, M., {et~al.} 2013{\natexlab{a}}, \apj, 764, 41

\bibitem[{Fragos {et~al.}(2013{\natexlab{b}})Fragos, Lehmer, Naoz, Zezas, \&
  Basu-Zych}]{Fragos2013a}
Fragos, T., Lehmer, B.~D., Naoz, S., Zezas, A., \& Basu-Zych, A.~R.
  2013{\natexlab{b}}, \apj, 776, L31

\bibitem[{Gabriel {et~al.}(2004)Gabriel, Denby, Fyfe, Hoar, Ibarra, Ojero,
  Osborne, Saxton, Lammers, \& Vacanti}]{gabriel2004}
Gabriel, C., Denby, M., Fyfe, D., {et~al.} 2004, in Astronomical Society of the
  Pacific Conference Series, Vol. 314, Astron. Data Anal. Softw. Syst. XIII,
  ed. D.~{Ochsenbein, F., Allen, M. G., Egret}, 759

\bibitem[{Giallongo {et~al.}(2015)Giallongo, Grazian, Fiore, Fontana,
  Pentericci, Vanzella, Dickinson, Kocevski, Castellano, Cristiani, Ferguson,
  Finkelstein, Grogin, Hathi, Koekemoer, Newman, \& Salvato}]{Giallongo2015}
Giallongo, E., Grazian, A., Fiore, F., {et~al.} 2015, \aap, 578, A83

\bibitem[{Gilfanov {et~al.}(2004{\natexlab{a}})Gilfanov, Grimm, \&
  Sunyaev}]{Gilfanov2004a}
Gilfanov, M., Grimm, H.~J., \& Sunyaev, R. 2004{\natexlab{a}}, \mnras, 347, L57

\bibitem[{Gilfanov {et~al.}(2004{\natexlab{b}})Gilfanov, Grimm, \&
  Sunyaev}]{Gilfanov2004b}
Gilfanov, M., Grimm, H.~J., \& Sunyaev, R. 2004{\natexlab{b}}, \mnras, 351,
  1365

\bibitem[{Gilfanov \& Merloni(2014)}]{Gilfanov2014}
Gilfanov, M. \& Merloni, A. 2014, Space Sci. Rev., 183, 121

\bibitem[{Grimm {et~al.}(2003)Grimm, Gilfanov, \& Sunyaev}]{Grimm2003}
Grimm, H.-J., Gilfanov, M., \& Sunyaev, R. 2003, \mnras, 339, 793

\bibitem[{Gunn \& Peterson(1965)}]{Gunn1965}
Gunn, J.~E. \& Peterson, B.~A. 1965, \apj, 142, 1633

\bibitem[{Haardt \& Salvaterra(2015)}]{Haardt2015}
Haardt, F. \& Salvaterra, R. 2015, \aap, 575, L16

\bibitem[{Hashimoto {et~al.}(2018)Hashimoto, Laporte, Mawatari, Ellis, Inoue,
  Zackrisson, Roberts-Borsani, Zheng, Tamura, Bauer, Fletcher, Harikane,
  Hatsukade, Hayatsu, Matsuda, Matsuo, Okamoto, Ouchi, Pell{\'{o}}, Rydberg,
  Shimizu, Taniguchi, Umehata, \& Yoshida}]{Hashimoto18}
Hashimoto, T., Laporte, N., Mawatari, K., {et~al.} 2018, \nat, 557, 392

\bibitem[{Hawley(2012)}]{Hawley2012}
Hawley, S.~A. 2012, Publ. Astron. Soc. Pacific, 124, 21

\bibitem[{Heckman {et~al.}(2011)Heckman, Borthakur, Overzier, Kauffmann,
  Basu-Zych, Leitherer, Sembach, Martin, Rich, Schiminovich, \&
  Seibert}]{Heckman2011}
Heckman, T.~M., Borthakur, S., Overzier, R., {et~al.} 2011, \apj, 730
  [\eprint[arXiv]{1101.4219}]

\bibitem[{Henry {et~al.}(2015)Henry, Scarlata, Martin, \& Erb}]{Henry15}
Henry, A., Scarlata, C., Martin, C., \& Erb, D. 2015, \apj, 809, 19

\bibitem[{Ho(2008)}]{Ho2008}
Ho, L.~C. 2008, Annu. Rev. \aap, 46, 475

\bibitem[{Hunter {et~al.}(2010)Hunter, Elmegreen, \& Ludka}]{Hunter2010}
Hunter, D.~A., Elmegreen, B.~G., \& Ludka, B.~C. 2010, Astron. J., 139, 447

\bibitem[{Iwasawa {et~al.}(2011)Iwasawa, Sanders, Teng, U, Armus, Evans,
  Howell, Komossa, Mazzarella, Petric, Surace, Vavilkin, Veilleux, \&
  Trentham}]{Iwasawa2011}
Iwasawa, K., Sanders, D.~B., Teng, S.~H., {et~al.} 2011, \aap, 529, A106

\bibitem[{Izotov {et~al.}(2011)Izotov, Guseva, \& Thuan}]{Izotov2011}
Izotov, Y., Guseva, N.~G., \& Thuan, T.~X. 2011, \apj, 728, 161

\bibitem[{Izotov {et~al.}(2016{\natexlab{a}})Izotov, Orlitova, Schaerer, Thuan,
  Verhamme, Guseva, \& Worseck}]{Izotov2016}
Izotov, Y., Orlitova, I., Schaerer, D., {et~al.} 2016{\natexlab{a}}, Nature,
  529, 178

\bibitem[{Izotov {et~al.}(2016{\natexlab{b}})Izotov, Schaerer, Thuan, Worseck,
  Guseva, Orlitova, \& Verhamme}]{Izotov2016b}
Izotov, Y., Schaerer, D., Thuan, T.~X., {et~al.} 2016{\natexlab{b}}, \mnras,
  461, 3683

\bibitem[{Izotov {et~al.}(2018{\natexlab{a}})Izotov, Schaerer, Worseck, Guseva,
  Thuan, Verhamme, Orlitov{\'{a}}, \& Fricke}]{Izotov2018a}
Izotov, Y., Schaerer, D., Worseck, G., {et~al.} 2018{\natexlab{a}}, \mnras,
  474, 4514

\bibitem[{Izotov {et~al.}(2018{\natexlab{b}})Izotov, Worseck, Schaerer, Guseva,
  Thuan, {Fricke A.}, \& Orlitov{\'{a}}}]{Izotov2018b}
Izotov, Y., Worseck, G., Schaerer, D., {et~al.} 2018{\natexlab{b}}, \mnras,
  478, 4851

\bibitem[{Jansen {et~al.}(2001)Jansen, Lumb, Altieri, Clavel, Ehle, Erd,
  Gabriel, Guainazzi, Gondoin, Much, Munoz, Santos, Schartel, Texier, \&
  Vacanti}]{2001A&A...365L...1J}
Jansen, F., Lumb, D., Altieri, B., {et~al.} 2001, \aap, 365, L1

\bibitem[{Jeon {et~al.}(2014)Jeon, Pawlik, Bromm, \&
  Milosavljevi{\'{c}}}]{Jeon2014}
Jeon, M., Pawlik, A.~H., Bromm, V., \& Milosavljevi{\'{c}}, M. 2014, \mnras,
  440, 3778

\bibitem[{Jia {et~al.}(2011)Jia, Ptak, Heckman, Overzier, Hornschemeier, \&
  LaMassa}]{Jia2011}
Jia, J., Ptak, A., Heckman, T.~M., {et~al.} 2011, \apj, 731, 55

\bibitem[{Justham \& Schawinski(2012)}]{Justham2012}
Justham, S. \& Schawinski, K. 2012, \mnras, 423, 1641

\bibitem[{Kaaret {et~al.}(2017{\natexlab{a}})Kaaret, Brorby, Casella, \&
  Prestwich}]{Kaaret2017}
Kaaret, P., Brorby, M., Casella, L., \& Prestwich, A.~H. 2017{\natexlab{a}},
  \mnras, 471, 4234

\bibitem[{Kaaret {et~al.}(2017{\natexlab{b}})Kaaret, Feng, \&
  Roberts}]{Kaaret2017a}
Kaaret, P., Feng, H., \& Roberts, T.~P. 2017{\natexlab{b}}, Annu. Rev. \aap,
  55, 303

\bibitem[{Kaaret {et~al.}(2011)Kaaret, Schmitt, \& Gorski}]{Kaaret2011}
Kaaret, P., Schmitt, J., \& Gorski, M. 2011, \apj, 741, 10

\bibitem[{Kalberla {et~al.}(2005)Kalberla, Burton, Hartmann, Arnal, Bajaja,
  Morras, \& Poppel}]{Kalberla2005}
Kalberla, P. M.~W., Burton, W.~B., Hartmann, D., {et~al.} 2005, \aap, 440, 9

\bibitem[{Kauffmann {et~al.}(2003)Kauffmann, Heckman, Tremonti, Brinchmann,
  Charlot, White, Ridgway, Brinkmann, Fukugita, Hall, Ivezi{\'{c}}, Richards,
  \& Schneider}]{Kauffmann2003}
Kauffmann, G., Heckman, T.~M., Tremonti, C., {et~al.} 2003, \mnras, 346, 1055

\bibitem[{Kennicutt(1998)}]{Kennicutt1998}
Kennicutt, R.~C. 1998, Annu. Rev. \aap, 36, 189

\bibitem[{Kewley \& Dopita(2002)}]{Kewley2002}
Kewley, L.~J. \& Dopita, M. 2002, \apjs, 142, 35

\bibitem[{Kewley {et~al.}(2013{\natexlab{a}})Kewley, Dopita, Leitherer,
  Dav{\'{e}}, Yuan, Allen, Groves, \& Sutherland}]{Kewley13}
Kewley, L.~J., Dopita, M., Leitherer, C., {et~al.} 2013{\natexlab{a}}, \apj,
  774, 100

\bibitem[{Kewley {et~al.}(2001)Kewley, Dopita, Sutherland, Heisler, \&
  Trevena}]{Kewley2001}
Kewley, L.~J., Dopita, M., Sutherland, R., Heisler, C., \& Trevena, J. 2001,
  \apj, 556, 121

\bibitem[{Kewley \& Ellison(2008)}]{Kewley2008}
Kewley, L.~J. \& Ellison, S.~L. 2008, \apj, 681, 1183

\bibitem[{Kewley {et~al.}(2013{\natexlab{b}})Kewley, Maier, Yabe, Ohta,
  Akiyama, Dopita, \& Yuan}]{Kewley13b}
Kewley, L.~J., Maier, C., Yabe, K., {et~al.} 2013{\natexlab{b}}, \apj, 774, L10

\bibitem[{Lehmer {et~al.}(2010)Lehmer, Alexander, Bauer, Brandt, Goulding,
  Jenkins, Ptak, \& Roberts}]{Lehmer2010}
Lehmer, B.~D., Alexander, D.~M., Bauer, F., {et~al.} 2010, \apj, 724, 559

\bibitem[{Leite {et~al.}(2017)Leite, Evoli, D'Angelo, Ciardi, Sigl, \&
  Ferrara}]{Leite2017}
Leite, N., Evoli, C., D'Angelo, M., {et~al.} 2017, \mnras, 469, 416

\bibitem[{Liu {et~al.}(2016)Liu, Slatyer, \& Zavala}]{Liu2016b}
Liu, H., Slatyer, T.~R., \& Zavala, J. 2016, Phys. Rev. D, 94, 063507

\bibitem[{Loaring {et~al.}(2005)Loaring, Dwelly, Page, Mason, McHardy, Gunn,
  Moss, Seymour, Newsam, Takata, Sekguchi, Sasseen, \& Cordova}]{Loaring2005}
Loaring, N.~S., Dwelly, T., Page, M.~J., {et~al.} 2005, \mnras, 362, 1371

\bibitem[{L{\'{o}}pez-S{\'{a}}nchez \& Esteban(2010)}]{Lopez-Sanchez2010}
L{\'{o}}pez-S{\'{a}}nchez, {\'{A}}.~R. \& Esteban, C. 2010, \aap, 517, A85

\bibitem[{Madau \& Fragos(2017)}]{Madau2017}
Madau, P. \& Fragos, T. 2017, \apj, 840, 39

\bibitem[{Madau \& Haardt(2015)}]{Madau2015}
Madau, P. \& Haardt, F. 2015, \apj, 813, L8

\bibitem[{Mapelli {et~al.}(2006)Mapelli, Ferrara, \& Pierpaoli}]{Mapelli2006}
Mapelli, M., Ferrara, A., \& Pierpaoli, E. 2006, \mnras, 369, 1719

\bibitem[{Mapelli {et~al.}(2010)Mapelli, Ripamonti, Zampieri, Colpi, \&
  Bressan}]{Mapelli2010}
Mapelli, M., Ripamonti, E., Zampieri, L., Colpi, M., \& Bressan, A. 2010,
  \mnras, 408, 234

\bibitem[{Marks {et~al.}(2012)Marks, Kroupa, Dabringhausen, \&
  Pawlowski}]{Marks2012}
Marks, M., Kroupa, P., Dabringhausen, J., \& Pawlowski, M.~S. 2012, \mnras,
  422, 2246

\bibitem[{Mateos {et~al.}(2008)Mateos, Warwick, Carrera, Stewart, Ebrero,
  {Della Ceca}, Caccianiga, Gilli, Page, Treister, Tedds, Watson, Lamer,
  Saxton, Brunner, \& Page}]{Mateos2008}
Mateos, S., Warwick, R.~S., Carrera, F.~J., {et~al.} 2008, \aap, 492, 51

\bibitem[{Mezcua {et~al.}(2018)Mezcua, Civano, Marchesi, Suh, Fabbiano, \&
  Volonteri}]{Mezcua2018}
Mezcua, M., Civano, F., Marchesi, S., {et~al.} 2018, \mnras, 478, 2576

\bibitem[{Mineo {et~al.}(2014)Mineo, Gilfanov, Lehmer, Morrison, \&
  Sunyaev}]{Mineo2014}
Mineo, S., Gilfanov, M., Lehmer, B.~D., Morrison, G.~E., \& Sunyaev, R. 2014,
  \mnras, 437, 1698

\bibitem[{Mineo {et~al.}(2012{\natexlab{a}})Mineo, Gilfanov, \&
  Sunyaev}]{Mineo2012}
Mineo, S., Gilfanov, M., \& Sunyaev, R. 2012{\natexlab{a}}, \mnras, 419, 2095

\bibitem[{Mineo {et~al.}(2012{\natexlab{b}})Mineo, Gilfanov, \&
  Sunyaev}]{Mineo2012b}
Mineo, S., Gilfanov, M., \& Sunyaev, R. 2012{\natexlab{b}}, \mnras, 426, 1870

\bibitem[{Mirabel {et~al.}(2011)Mirabel, Dijkstra, Laurent, Loeb, \&
  Pritchard}]{Mirabel2011}
Mirabel, I.~F., Dijkstra, M., Laurent, P., Loeb, A., \& Pritchard, J.~R. 2011,
  \aap, 528, A149

\bibitem[{Mullaney \& Ward(2008)}]{Mullaney08}
Mullaney, J. \& Ward, M. 2008, \mnras, 385, 53

\bibitem[{Nath \& Biermann(1993)}]{Nath1993}
Nath, B.~B. \& Biermann, P.~L. 1993, \mnras, 265, 241

\bibitem[{Oesch {et~al.}(2018)Oesch, Bouwens, Illingworth, Labb{\'{e}}, \&
  Stefanon}]{Oesch18}
Oesch, P., Bouwens, R., Illingworth, G., Labb{\'{e}}, I., \& Stefanon, M. 2018,
  \apj, 855, 105

\bibitem[{Orlitov{\'{a}} {et~al.}(2018)Orlitov{\'{a}}, Verhamme, Henry,
  Scarlata, Jaskot, Oey, \& Schaerer}]{Orlitova2018}
Orlitov{\'{a}}, I., Verhamme, A., Henry, A., {et~al.} 2018, \aap, 616, A60

\bibitem[{Ot{\'{i}}-Floranes {et~al.}(2012)Ot{\'{i}}-Floranes, Mas-Hesse,
  Jim{\'{e}}nez-Bail{\'{o}}n, Schaerer, Hayes, {\"{O}}stlin, Atek, \&
  Kunth}]{Oti-Floranes2012}
Ot{\'{i}}-Floranes, H., Mas-Hesse, J.~M., Jim{\'{e}}nez-Bail{\'{o}}n, E.,
  {et~al.} 2012, \aap, 546, A65

\bibitem[{Ot{\'{i}}-Floranes {et~al.}(2014)Ot{\'{i}}-Floranes, Mas-Hesse,
  Jim{\'{e}}nez-Bail{\'{o}}n, Schaerer, Hayes, {\"{O}}stlin, Atek, \&
  Kunth}]{Oti-Floranes2014}
Ot{\'{i}}-Floranes, H., Mas-Hesse, J.~M., Jim{\'{e}}nez-Bail{\'{o}}n, E.,
  {et~al.} 2014, \aap, 566, A38

\bibitem[{Ouchi {et~al.}(2018)Ouchi, Harikane, Shibuya, Shimasaku, Taniguchi,
  Konno, Kobayashi, Kajisawa, Nagao, Ono, Inoue, Umemura, Mori, Hasegawa,
  Higuchi, Komiyama, Matsuda, Nakajima, Saito, \& Wang}]{Ouchi18}
Ouchi, M., Harikane, Y., Shibuya, T., {et~al.} 2018, Publ. Astron. Soc. Japan,
  70, S13

\bibitem[{Ouchi {et~al.}(2009)Ouchi, Mobasher, Shimasaku, Ferguson, Fall, Ono,
  Kashikawa, Morokuma, Nakajima, Okamura, Dickinson, Giavalisco, \&
  Ohta}]{Ouchi09b}
Ouchi, M., Mobasher, B., Shimasaku, K., {et~al.} 2009, \apj, 706, 1136

\bibitem[{Overzier {et~al.}(2009)Overzier, Heckman, Tremonti, Armus, Basu-Zych,
  Gon{\c{c}}alves, Rich, Martin, Ptak, Schiminovich, Ford, Madore, \&
  Seibert}]{Overzier09}
Overzier, R.~A., Heckman, T.~M., Tremonti, C., {et~al.} 2009, \apj, 706, 203

\bibitem[{Paardekooper {et~al.}(2015)Paardekooper, Khochfar, \&
  Vecchia}]{Paardekooper2015}
Paardekooper, J.~P., Khochfar, S., \& Vecchia, C.~D. 2015, \mnras, 451, 2544

\bibitem[{Pettini \& Pagel(2004)}]{Pettini2004}
Pettini, M. \& Pagel, B. E.~J. 2004, \mnras, 348, L59

\bibitem[{Prestwich {et~al.}(2015)Prestwich, Jackson, Kaaret, Brorby, Roberts,
  Saar, \& Yukita}]{Prestwich2015}
Prestwich, A.~H., Jackson, F., Kaaret, P., {et~al.} 2015, \apj, 812, 166

\bibitem[{Puschnig {et~al.}(2017)Puschnig, Hayes, {\"{O}}stlin, Rivera-Thorsen,
  Melinder, Cannon, Menacho, Zackrisson, Bergvall, \& Leitet}]{Puschnig2017}
Puschnig, J., Hayes, M., {\"{O}}stlin, G., {et~al.} 2017, \mnras, 469, 3252

\bibitem[{Ranalli {et~al.}(2003)Ranalli, Comastri, \& Setti}]{Ranalli2003}
Ranalli, P., Comastri, A., \& Setti, G. 2003, \aap, 399, 39

\bibitem[{Robertson {et~al.}(2010)Robertson, Ellis, Dunlop, McLure, \&
  Stark}]{Robertson2010}
Robertson, B.~E., Ellis, R.~S., Dunlop, J.~S., McLure, R.~J., \& Stark, D.~P.
  2010, Nature, 468, 49

\bibitem[{Robertson {et~al.}(2015)Robertson, Ellis, Furlanetto, \&
  Dunlop}]{Robertson2015}
Robertson, B.~E., Ellis, R.~S., Furlanetto, S.~R., \& Dunlop, J.~S. 2015, \apj,
  802, L19

\bibitem[{Rosen {et~al.}(2016)Rosen, Webb, Watson, Ballet, Barret, Braito,
  Carrera, Ceballos, Ceca, Denkinson, Esquej, Farrell, Freyberg, Guillout,
  Heil, Lamer, Lin, Martino, Michel, Motch, Gomez-moran, Page, Page, Pakull,
  Pye, Read, Rodriguez, Sakano, Saxton, Schwope, Sturm, Traulsen, Yershov, \&
  Zolotukhin}]{Rosen2016}
Rosen, S.~R., Webb, N.~A., Watson, M.~G., {et~al.} 2016, \aap, 590, 1

\bibitem[{Sazonov \& Khabibullin(2018)}]{Sazonov2018}
Sazonov, S. \& Khabibullin, I. 2018, \mnras, 476, 2530

\bibitem[{Sazonov \& Sunyaev(2015)}]{Sazonov2015}
Sazonov, S. \& Sunyaev, R. 2015, \mnras, 454, 3464

\bibitem[{Schaerer {et~al.}(2016)Schaerer, Izotov, Verhamme, Orlitov{\'{a}},
  Thuan, Worseck, \& Guseva}]{Schaerer16}
Schaerer, D., Izotov, Y., Verhamme, A., {et~al.} 2016, \aap, 591, L8

\bibitem[{Schenker {et~al.}(2013)Schenker, Robertson, Ellis, Ono, McLure,
  Dunlop, Koekemoer, Bowler, Ouchi, Curtis-Lake, Rogers, Schneider, Charlot,
  Stark, Furlanetto, \& Cirasuolo}]{Schenker2013}
Schenker, M.~A., Robertson, B.~E., Ellis, R.~S., {et~al.} 2013, \apj, 768, 196

\bibitem[{Shapley {et~al.}(2016)Shapley, Steidel, Strom, Bogosavljevi{\'{c}},
  Reddy, Siana, Mostardi, \& Rudie}]{Shapley2016}
Shapley, A.~E., Steidel, C.~C., Strom, A.~L., {et~al.} 2016, \apj, 826, L24

\bibitem[{Smith {et~al.}(2006)Smith, Westmoquette, Gallagher, O'Connell,
  Rosario, \& {De Grijs}}]{Smith2006a}
Smith, L.~J., Westmoquette, M.~S., Gallagher, J.~S., {et~al.} 2006, \mnras,
  370, 513

\bibitem[{Smith \& Guainazzi(2016)}]{Smith2016}
Smith, M. \& Guainazzi, M. 2016, XMM-SOC-CAL-TN-0018

\bibitem[{Snijders {et~al.}(2007)Snijders, Kewley, \& van~der
  Werf}]{Snijders2007}
Snijders, L., Kewley, L.~J., \& van~der Werf, P.~P. 2007, \apj, 669, 269

\bibitem[{Stanway {et~al.}(2016)Stanway, Eldridge, \& Becker}]{Stanway2016}
Stanway, E.~R., Eldridge, J.~J., \& Becker, G.~D. 2016, \mnras, 456, 485

\bibitem[{Stark(2016)}]{Stark2016}
Stark, D.~P. 2016, Annu. Rev. \aap, 54, 761

\bibitem[{Str{\"{u}}der {et~al.}(2001)Str{\"{u}}der, Briel, Dennerl, Hartmann,
  Kendziorra, Meidinger, Pfeffermann, Reppin, Aschenbach, Bornemann,
  Br{\"{a}}uninger, Burkert, Elender, Freyberg, Haberl, Hartner, Heuschmann,
  Hippmann, Kastelic, Kemmer, Kettenring, Kink, Krause, M{\"{u}}ller, Oppitz,
  Pietsch, Popp, Predehl, Read, Stephan, St{\"{o}}tter, Tr{\"{u}}mper, Holl,
  Kemmer, Soltau, St{\"{o}}tter, Weber, Weichert, von Zanthier, Carathanassis,
  Lutz, Richter, Solc, B{\"{o}}ttcher, Kuster, Staubert, Abbey, Holland,
  Turner, Balasini, Bignami, {La Palombara}, Villa, Buttler, Gianini,
  Lain{\'{e}}, Lumb, \& Dhez}]{pn}
Str{\"{u}}der, L., Briel, U., Dennerl, K., {et~al.} 2001, \aap, 365, L18

\bibitem[{Svoboda {et~al.}(2017)Svoboda, Guainazzi, \& Merloni}]{Svoboda2017}
Svoboda, J., Guainazzi, M., \& Merloni, A. 2017, \aap, 603, A127

\bibitem[{Tremonti {et~al.}(2004)Tremonti, Heckman, Kauffmann, Brinchmann,
  Charlot, White, Seibert, Peng, Schlegel, Uomoto, Fukugita, \&
  Brinkmann}]{Tremonti2004}
Tremonti, C.~A., Heckman, T.~M., Kauffmann, G., {et~al.} 2004, \apj, 613, 898

\bibitem[{Tueros {et~al.}(2014)Tueros, del Valle, \& Romero}]{Tueros2014}
Tueros, M., del Valle, M.~V., \& Romero, G.~E. 2014, \aap, 570, L3

\bibitem[{Turner {et~al.}(2001)Turner, Abbey, Arnaud, Balasini, Barbera,
  Belsole, Bennie, Bernard, Bignami, Boer, Briel, Butler, Cara, Chabaud, Cole,
  Collura, Conte, Cros, Denby, Dhez, {Di Coco}, Dowson, Ferrando, Ghizzardi,
  Gianotti, Goodall, Gretton, Griffiths, Hainaut, Hochedez, Holland, Jourdain,
  Kendziorra, Lagostina, Laine, {La Palombara}, Lortholary, Lumb, Marty,
  Molendi, Pigot, Poindron, Pounds, Reeves, Reppin, Rothenflug, Salvetat,
  Sauvageot, Schmitt, Sembay, Short, Spragg, Stephen, Str�der, Tiengo,
  Trifoglio, Tr{\"{u}}mper, Vercellone, Vigroux, Villa, Ward, Whitehead, \&
  Zonca}]{mos}
Turner, M. J.~L., Abbey, A., Arnaud, M., {et~al.} 2001, \aap, 365, L27

\bibitem[{Vald{\'{e}}s {et~al.}(2007)Vald{\'{e}}s, Ferrara, Mapelli, \&
  Ripamonti}]{Valdes2007}
Vald{\'{e}}s, M., Ferrara, A., Mapelli, M., \& Ripamonti, E. 2007, \mnras, 377,
  245

\bibitem[{Valtchanov {et~al.}(2001)Valtchanov, Pierre, \&
  Gastaud}]{Valtchanov2001}
Valtchanov, I., Pierre, M., \& Gastaud, R. 2001, \aap, 370, 689

\bibitem[{Vanzella {et~al.}(2018)Vanzella, Nonino, Cupani, Castellano, Sani,
  Mignoli, Calura, Meneghetti, Gilli, Comastri, Mercurio, Caminha, Caputi,
  Rosati, Grillo, Cristiani, Balestra, Fontana, \& Giavalisco}]{Vanzella18}
Vanzella, E., Nonino, M., Cupani, G., {et~al.} 2018, \mnras, 476, L15

\bibitem[{Verhamme {et~al.}(2017)Verhamme, Orlitov{\'{a}}, Schaerer, Izotov,
  Worseck, Thuan, \& Guseva}]{Verhamme2017}
Verhamme, A., Orlitov{\'{a}}, I., Schaerer, D., {et~al.} 2017, \aap, 597, A13

\bibitem[{Watson {et~al.}(2009)Watson, Schr{\"{o}}der, Fyfe, Page, Lamer,
  Mateos, Pye, Sakano, Rosen, Ballet, Barcons, Barret, Boller, Brunner, Brusa,
  Caccianiga, Carrera, Ceballos, {Della Ceca}, Denby, Denkinson, Dupuy,
  Farrell, Fraschetti, Freyberg, Guillout, Hambaryan, Maccacaro, Mathiesen,
  McMahon, Michel, Motch, Osborne, Page, Pakull, Pietsch, Saxton, Schwope,
  Severgnini, Simpson, Sironi, Stewart, Stewart, Stobbart, Tedds, Warwick,
  Webb, West, Worrall, \& Yuan}]{Watson2009}
Watson, M., Schr{\"{o}}der, A., Fyfe, D., {et~al.} 2009, \aap, 493, 339

\bibitem[{Xu {et~al.}(2014)Xu, Ahn, Wise, Norman, \& O'Shea}]{Xu2014}
Xu, H., Ahn, K., Wise, J.~H., Norman, M.~L., \& O'Shea, B.~W. 2014, \apj, 791,
  110

\bibitem[{Yajima {et~al.}(2009)Yajima, Umemura, Mori, \& Nakamoto}]{Yajima09}
Yajima, H., Umemura, M., Mori, M., \& Nakamoto, T. 2009, \mnras, 398, 715

\bibitem[{Zaroubi(2012)}]{Zaroubi2012}
Zaroubi, S. 2012, ArXiv e-prints [\eprint[arXiv]{1206.0267}]

\end{thebibliography}






\appendix
\section{Details of the Source and Background Region Extractions}
\label{Details_extraction}

The source and background extraction regions for all three GPs studied in this paper are shown in Figs.~\ref{image_gp_1}-\ref{image_gp_3}.
The coordinates and sizes of these regions are summarized in Table~\ref{extraction_regions}.

\smallskip

\begin{table*}[tbh]

\caption{Source and Background Extraction Regions.}
\centering
\begin{tabular}{ccccc}
 \rule{0cm}{0.5cm}
 Source & Detector  & Source Extraction Region & Background Extraction Region \\
 \hline \hline
 \rule[-0.7em]{0pt}{2em} 
SDSSJ074936.7+333716 (GP1) &  PN &  7:49:36.770, +33:37:16.30, 24\arcsec  &  7:49:42.062, +33:36:57.15, 30\arcsec   \\
 \rule[-0.7em]{0pt}{2em} 
 &   &   & + 7:49:28.879, +33:37:27.82, 30\arcsec   \\
 \rule[-0.7em]{0pt}{2em}   
 &  MOS &  7:49:36.770, +33:37:16.30, 24\arcsec &   7:49:47.665, +33:37:39.02, 90\arcsec  \\
 \rule[-0.7em]{0pt}{2em}   
SDSSJ082247.6+224144 (GP2) &  PN &  8:22:47.660, +22:41:44.00, 30\arcsec &  8:22:42.396, +22:42:02.20, 30\arcsec  \\
 \rule[-0.7em]{0pt}{2em}   
 &  MOS &  8:22:47.660, +22:41:44.00, 30\arcsec &   8:22:47.776, +22:38:20.88, 90\arcsec  \\
 \rule[-0.7em]{0pt}{2em} 
SDSSJ133928.3+151642 (GP3) &  PN &  13:39:28.300, +15:16:42.10, 30\arcsec  &  13:39:34.794, +15:16:48.43, 60\arcsec     \\
 \rule[-0.7em]{0pt}{2em}   
 &  MOS &  13:39:28.300, +15:16:42.10, 30\arcsec &   13:39:35.287, +15:14:29.70, 120\arcsec  \\
\end{tabular}
\tablecomments{
The listed coordinates represent the FK5 system sexagesimal coordinates (J2000)
of the centers of circular source/background extraction regions and the circle radii in arcsec. 
Two circular regions were combined to have a larger extraction area 
for the background in the first PN observation.
}
\label{extraction_regions} 
\end{table*}

\end{document}